\documentclass[%
superscriptaddress,
 amsmath,amssymb,
 aps, prl, twocolumn,
]{revtex4-2}

\usepackage{graphicx}
\usepackage{dcolumn}
\usepackage{bm}
\usepackage[unicode=true,bookmarks=true,bookmarksnumbered=false,bookmarksopen=false,breaklinks=false,pdfborder={0 0 1},backref=false,colorlinks=true]{hyperref}

\usepackage{physics, mathtools, qcircuit}
\usepackage[normalem]{ulem}
\hypersetup{linkcolor=magenta, urlcolor=blue, citecolor=blue, pdfstartview={FitH}}
\newcommand{\ha}{\hat{a}}
\newcommand{\hadag}{\hat{a}^\dagger}
\newcommand{\hn}{\hat{n}}
\newcommand{\rmd}{\mathrm{d}}

\begin{document}

\preprint{APS/123-QED}

\title{Echoed Random Quantum Metrology}

\affiliation{Laboratory of Quantum Information, University of Science and Technology of China, Hefei 230026, China}
\affiliation{Center for Quantum Information, Institute for Interdisciplinary Information Sciences, Tsinghua University, Beijing 100084, China}
\affiliation{Anhui Province Key Laboratory of Quantum Network, University of Science and Technology of China, Hefei 230026, China}
\affiliation{Center For Excellence in Quantum Information and Quantum Physics, University of Science and Technology of China, Hefei 230026, China}
\affiliation{Hefei National Laboratory, Hefei 230088, China}

\author{Dong-Sheng Liu}
\thanks{These authors contributed equally.}
\affiliation{Laboratory of Quantum Information, University of Science and Technology of China, Hefei 230026, China}
\affiliation{Anhui Province Key Laboratory of Quantum Network, University of Science and Technology of China, Hefei 230026, China}
\affiliation{Hefei National Laboratory, Hefei 230088, China}

\author{Zi-Jie Chen}
\thanks{These authors contributed equally.}
\affiliation{Laboratory of Quantum Information, University of Science and Technology of China, Hefei 230026, China}
\affiliation{Anhui Province Key Laboratory of Quantum Network, University of Science and Technology of China, Hefei 230026, China}

\author{Ziyue Hua}
\affiliation{Center for Quantum Information, Institute for Interdisciplinary Information Sciences, Tsinghua University, Beijing 100084, China}

\author{Yilong Zhou}
\affiliation{Center for Quantum Information, Institute for Interdisciplinary Information Sciences, Tsinghua University, Beijing 100084, China}

\author{Qing-Xuan Jie}
\affiliation{Laboratory of Quantum Information, University of Science and Technology of China, Hefei 230026, China}
\affiliation{Anhui Province Key Laboratory of Quantum Network, University of Science and Technology of China, Hefei 230026, China}

\author{Weizhou Cai}
\affiliation{Laboratory of Quantum Information, University of Science and Technology of China, Hefei 230026, China}
\affiliation{Anhui Province Key Laboratory of Quantum Network, University of Science and Technology of China, Hefei 230026, China}

\author{Ming Li}
\affiliation{Laboratory of Quantum Information, University of Science and Technology of China, Hefei 230026, China}
\affiliation{Anhui Province Key Laboratory of Quantum Network, University of Science and Technology of China, Hefei 230026, China}

\author{Luyan Sun}
\email{luyansun@tsinghua.edu.cn}
\affiliation{Center for Quantum Information, Institute for Interdisciplinary Information Sciences, Tsinghua University, Beijing 100084, China}
\affiliation{Hefei National Laboratory, Hefei 230088, China}

\author{Chang-Ling Zou}
\email{clzou321@ustc.edu.cn}
\affiliation{Laboratory of Quantum Information, University of Science and Technology of China, Hefei 230026, China}
\affiliation{Anhui Province Key Laboratory of Quantum Network, University of Science and Technology of China, Hefei 230026, China}
\affiliation{Center For Excellence in Quantum Information and Quantum Physics, University of Science and Technology of China, Hefei 230026, China}
\affiliation{Hefei National Laboratory, Hefei 230088, China}
\affiliation{Shenzhen International Quantum Academy, Shenzhen 518048, China}

\author{Xi-Feng Ren}
\email{renxf@ustc.edu.cn}
\affiliation{Laboratory of Quantum Information, University of Science and Technology of China, Hefei 230026, China}
\affiliation{Anhui Province Key Laboratory of Quantum Network, University of Science and Technology of China, Hefei 230026, China}
\affiliation{Center For Excellence in Quantum Information and Quantum Physics, University of Science and Technology of China, Hefei 230026, China}

\affiliation{Hefei National Laboratory, Hefei 230088, China}

\author{Guang-Can Guo}
\affiliation{Laboratory of Quantum Information, University of Science and Technology of China, Hefei 230026, China}
\affiliation{Anhui Province Key Laboratory of Quantum Network, University of Science and Technology of China, Hefei 230026, China}
\affiliation{Center For Excellence in Quantum Information and Quantum Physics, University of Science and Technology of China, Hefei 230026, China}

\affiliation{Hefei National Laboratory, Hefei 230088, China}

\date{\today}
\begin{abstract}
Quantum metrology typically demands the preparation of exotic quantum probe states, such as entangled or squeezed states, to surpass classical limits. However, the need for carefully calibrated system parameters and finely optimized quantum controls imposes limitations on scalability and robustness. Here, we circumvent these limitations by introducing an echoed random process that achieves sensitivity approaching the Heisenberg limit while remaining blind to the random probe state. We demonstrate that by simply driving a Kerr nonlinear mode with random pulses, the emergence of sub-Planck phase-space structures grants high sensitivity, eliminating the need for complex quantum control. The protocol is statistically robust, yielding high performance across broad driving parameter ranges while exhibiting resilience to control fluctuations and photon loss. Broadly applicable to both bosonic and qubit platforms, our work reveals a practical, hardware-efficient, scalable, and optimization-free route to quantum-enhanced metrology in high-dimensional Hilbert spaces.
\end{abstract}

\maketitle

\textit{\textbf{Introduction}.-} Quantum metrology leverages quantum effects to enhance measurement precision beyond the standard quantum limit (SQL), potentially reaching the fundamental Heisenberg limit (HL)~\cite{Giovannetti2011,Degen2017}. The realization of this enhancement relies on the use of exotic quantum probe states, such as Greenberger-Horne-Zeilinger (GHZ) states~\cite{Bao2024}, Fock states~\cite{deng2024,wolf2019}, spin-squeezed states~\cite{Eckner2023}, and maximum-variance states~\cite{Wang2019b}. However, the preparation of these states is highly demanding, requiring precise calibration of system parameters, high-fidelity control over the quantum dynamics, and computationally expensive classical optimization. These stringent requirements on hardware performance and classical computational resources have largely confined such demonstrations to small-scale systems, thus highlighting a significant challenge in their scalability~\cite{Pan2025}. Consequently, implementing scalable and robust metrology in the high-dimensional Hilbert space remains an open problem.

Remarkably, the strict requirement for fine-tuned control is being challenged by emerging paradigms like quantum reservoir computing ~\cite{Mujal2021,Hu2024,Wang2025a}, where unengineered dynamical evolutions of quantum systems are successfully exploited for information processing. Parallel efforts in metrology have similarly indicated that randomized states and measurements can yield quantum enhancement ~\cite{Oszmaniec2016, Shi2025, Kobrin2024, Zhu2018, Hou2018, Zhou2025}. However, these approaches often merely shift the complexity from state preparation to detection, necessitating experimentally demanding state-dependent or collective measurements. Such overhead can be circumvented through the echo protocols~\cite{Davis2016, Linnemann2016, Nolan2017, Haine2018, Mirkhalaf2018, Lewis-Swan2020, Gilmore2021, Colombo2022, Li2023a, Guo2024, Abanin2025}, where a quantum system evolves under a unitary $U$ into a highly sensitive nonclassical state. Following the imprinting of a perturbation, the time-reversed evolution $U^\dagger$ is applied to map the signal back into a simple observable, such as the initial state population.

Here, we propose and analyze an echoed random quantum metrology scheme by combining unengineered dynamical evolutions with the echo technique. Random probe states are generated by stochastic dynamical evolution in a nonlinear system, eliminating the need for delicate state preparation and measurement. Time-reversed dynamics followed by single-photon measurement is used to extract the unknown parameters, thereby removing the need for measurement optimization. Notably, in our scheme, one does not need to know the exact form of the probe state. This fully optimization-free paradigm inherently boosts scalability towards a higher-dimensional Hilbert space. Strikingly, this method achieves near-Heisenberg-limit precision with high probabilities over broad driving parameter ranges. The resulting metrological gain is resilient to both control fluctuations and decoherence. Although not aiming to outperform the best-optimized schemes, our results demonstrate a hardware-efficient, robust, and scalable alternative to quantum-enhanced sensing by harnessing, rather than avoiding, complexity and randomness in nonlinear systems.

\begin{figure}
    \centering
    \includegraphics[width=\linewidth]{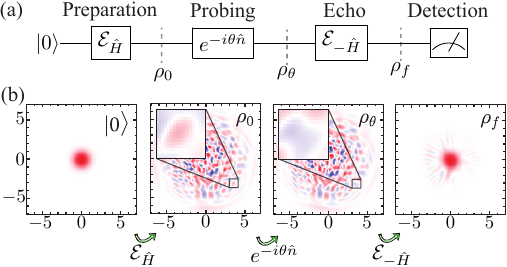}
    \caption{(a) Schematic illustration of the phase estimation procedure using the echoed random quantum metrology protocol. Probe states are prepared via dynamics $\mathcal{E}_{\hat{H}}$, driven by randomly modulated coherent pulses. A time-reversed evolution $\mathcal{E}_{-\hat{H}}$ is applied before measurement, effectively realizing a non-Gaussian measurement that accesses information stored in higher-order moments. (b) Evolution of the Wigner function, starting from the vacuum state, under the forward random channel, probing, and the corresponding time-reversed channel. The phase $\theta$ induces a rotation in phase space, and the fine structures of the distributions make the states highly sensitive to these rotations. As the magnified insets demonstrate, even a small rotation in phase space induces a sign flip in the Wigner function.}
    \label{fig1}
\end{figure}

\textit{\textbf{Echoed random quantum metrology}.-} Figure~\ref{fig1}(a) illustrates the protocol, where we consider a phase estimation protocol using a single bosonic mode with a self-Kerr nonlinearity. The system is initialized in the vacuum state $\ket{0}$, and then proceeds through four stages, dynamical evolution for random state preparation ($\mathcal{E}_{  \hat{H}}$), probing, the reversed evolution as echo  ($\mathcal{E}_{-\hat{H}}$), and detection. The system evolution is governed by the time-dependent Hamiltonian ($\hbar=1$):
\begin{equation}\label{eq:hamiltonian}
    \hat{H}(t) = \chi(\hadag\ha)^2 + u_1(t)(\hadag+\ha) + iu_2(t)(\hadag-\ha),
\end{equation}
where $\chi$ is the Kerr nonlinearity strength and $u_{1,2}(t)$ are the coherent driving amplitudes. Crucially, the drive pulses are not optimized but rather chosen randomly: random amplitudes are step-wise constant with step size $\tau$, and their strengths are bounded within $[-\epsilon,\epsilon]$. This evolution for a duration $T$ defines a random encoding map $\mathcal{E}_{\hat{H}}$, which generates the probe state $\rho_0 = \mathcal{E}_{\hat{H}}(\dyad{0})$. During the probing stage, the estimated parameter $\theta$ is imprinted onto the state, yielding $\rho_\theta = e^{-i\theta \hn} \rho_0 e^{i\theta \hn}$, where $\hn = \hadag \ha$.  Then, an echoed process with the Hamiltonian $ \hat{H}^\prime(t)=- \hat{H}(T-t)$ refocuses the mode state back towards the vacuum state, effectively mapping the encoded information onto a low-photon-number subspace, which can be effectively detected through photon-number detection eventually.

Figure~\ref{fig1}(b) visualizes the Wigner function evolution through the protocol. The interplay between random driving and the self-Kerr nonlinearity is sufficient to generate intricate sub-Planck-scale structures within the phase space~\cite{Zurek2001, Toscano2006}. Consequently, the resulting probe state becomes exceptionally sensitive to rotations in this space, and this is the origin of the quantum enhancement. Since no classical optimization is needed in the process, this protocol is naturally scalable to a high-dimensional Hilbert space.

This bosonic mode system with Kerr nonlinearity is general and can be realized in diverse physical platforms, including superconducting circuits~\cite{deng2024,Wang2022a} and trapped ions~\cite{Nikolova2025,wolf2019}. In the following, we focus on the superconducting circuit system, where  $\chi$ can be continuously tuned and even sign-inverted through the integration of a superconducting nonlinear asymmetric inductive element (SNAIL)~\cite{He2023} or by engineering the cavity-transmon coupling~\cite{Xu2025}. We note that this method of using a time-reversed interaction for signal amplification is also known as the SATIN protocol~\cite{Colombo2022, Li2023a} or interaction-based readouts~\cite{Davis2016, Nolan2017, Haine2018, Mirkhalaf2018}.

\textit{\textbf{Statistical robustness}.-} A primary advantage of the echoed random protocol is its statistical robustness: a significant metrological gain is achievable for the vast majority of random probe states, ensuring high sensitivity without the need for optimization. The performance of our protocol can be quantitatively evaluated through the classical Fisher information (CFI) according to the Cram\'er-Rao bound~\cite{Yuan2017}. For a fixed measurement scheme described by the POVM elements $\{\hat{M}_n\}$, the attainable precision of the estimated parameter $\theta$ is fundamentally determined by the CFI
\begin{equation}\label{eq:CFI}
    I_\mathrm{c}(\theta_0) = \left. \sum_n\frac{1}{p_n( \theta)}\left[\frac{\rmd p_n( \theta)}{\rmd\theta}\right]^2\right|_{\theta=\theta_0}.
\end{equation}
Here,  $p_n( \theta) = \tr(\hat{M}_n\rho_f)$ is the probability of measurement outcome $n$, and $\theta_0$ is the bias phase at the probing stage to enhance the sensitivity ${\rmd p_n( \theta)}/{\rmd\theta}$. Considering the capability of the photon-number detection realized in experimental systems~\cite{Heeres2015,Cai2021,Meekhof1996,Li2025}, we choose the POVM elements $\{\hat{M}_0=\dyad{0}, \hat{M}_1=I-\dyad{0}\}$ for single-photon detection in the following.

For a coherent state with the same average photon number $\ev{\hn}=\tr(\hn\rho_0)$, the CFI is upper bounded by $4\ev{\hn}$ (see Supplementary Material). Therefore, the metrological gain of our protocol compared to this classical resource is
\begin{equation}\label{eq:Gain}
    G_\mathrm{c}(\theta_0) \coloneq \frac{I_c(\theta_0)}{4\ev{\hn}}.
\end{equation}
The maximum CFI and metrological gain over all bias points $\theta_0$ are defined as $I_\mathrm{c,max} \coloneq \max_{\theta_0} I_\mathrm{c}(\theta_0)$ and $G_\mathrm{c,max} \coloneq \max_{\theta_0} G_\mathrm{c}(\theta_0)$, respectively, with the corresponding optimal bias point denoted by $\theta_\mathrm{b}$.


\begin{figure*}
    \centering
    \includegraphics[width=\linewidth]{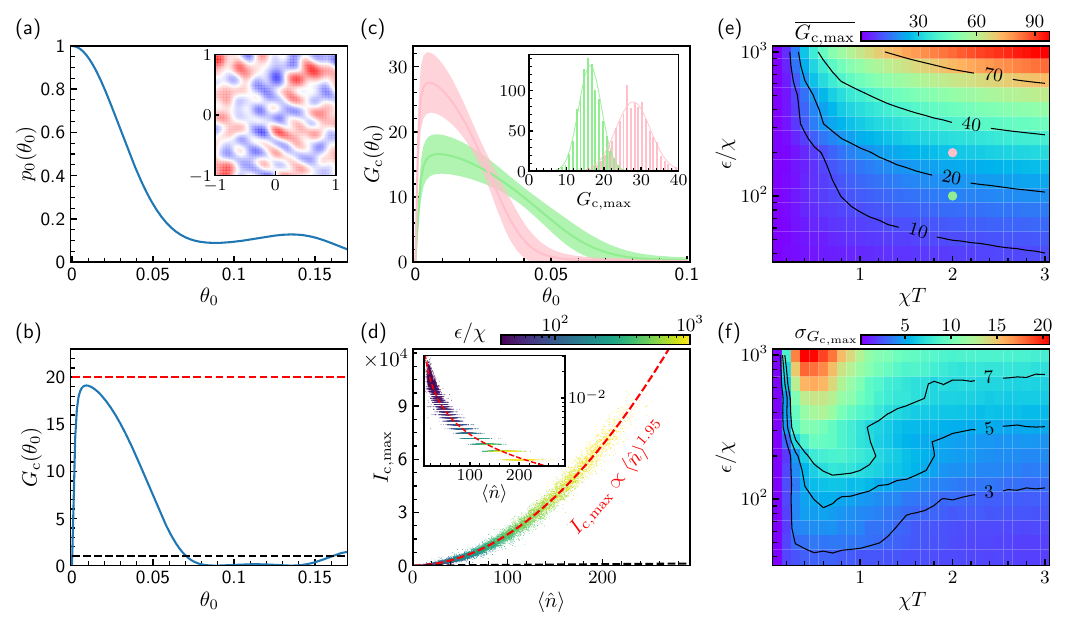}
    \caption{(a) Probability $p_0(\theta_0)$ versus the bias point $\theta_0$ for a randomly generated probe state with $\epsilon/\chi = 100, \chi T = 2$. Inset: corresponding Wigner function with average photon number $\ev{\hat{n}}\approx 27$.
    (b) Corresponding metrological gain $G_\mathrm{c}(\theta_0)$ of (a), where the black dashed line marks the standard quantum limit (SQL), $G_\mathrm{SQL}=1$, and the red dashed line indicates the quantum Fisher information bound.
    (c) $G_\mathrm{c}(\theta_0)$ for $\epsilon/\chi = 100$ (light green) and $\epsilon/\chi = 200$ (pink), both at $\chi T = 2$. Shaded areas denote standard deviations over the realizations. Inset: corresponding histograms of $G_\mathrm{c,max}$ at the optimal bias point, overlaid with Gaussian fits.
    (d) Scatter plot of classical Fisher information (CFI) $I_\mathrm{c,max}$ versus $\ev{\hn}$ at the optimal bias point for $\chi T=2$. The black dashed line marks the SQL $I_\mathrm{SQL}=4\ev{\hn}$ while the red dashed curve shows a power-law fit, $I_\mathrm{c,max}=2.17\times\ev{\hn}^{1.95}$, approaching the Heisenberg scaling $I_\mathrm{HL}\propto\ev{\hn}^2$. Inset: corresponding optimal bias point $\theta_\mathrm{b}$ versus $\ev{\hn}$, with weighted power-law fit: $\theta_\mathrm{b}=0.23\times \ev{\hn}^{-0.89}$.
    (e) The mean of $G_\mathrm{c,max}$ over 1000 realizations, denoted as $\overline{G_\mathrm{c,max}}$. The light green and pink markers indicate the parameter sets $(\epsilon/\chi=100, \chi T=2)$ and $(\epsilon/\chi=200, \chi T=2)$ used in (c).
    (f) The standard deviation of $G_\mathrm{c,max}$ over 1000 realizations, denoted as $\sigma_{G_\mathrm{c,max}}$. The system exhibits a transient regime at early times, entering a stabilization phase for $\chi T \gtrsim 1.5$ where the variance remains low. Contour lines (black) are included in (e–f).
    Other parameters: $\tau = 0.1/\chi$, $d = 1050$, and $\epsilon_\mathrm{dp} = 10^{-3}$.}
    \label{fig2}
\end{figure*}

We first consider the ideal case without decoherence during the forward and backward evolution, i.e., both $\mathcal{E}_{\hat{H}}$ and $\mathcal{E}_{-\hat{H}}$ are unitary channels, and we assume single-photon detection. We truncate the Fock space to dimension $d$ and apply a depolarizing channel with strength $\epsilon_\mathrm{dp}$ to the final state $\rho_f$, modeling measurement noise. Figure \ref{fig2}(a) shows the probability $p_0(\theta_0)$ under different bias points $\theta_0$ for a representative randomly generated probe state, along with its corresponding phase-space distribution. The intricate structures in phase space contribute to the high sensitivity to small phase shifts. The corresponding metrological gain obtained from Eqs.~\eqref{eq:CFI} and \eqref{eq:Gain} is plotted in Fig. \ref{fig2}(b). It is shown that the metrological gain of this probe state closely approaches the maximum achievable value at the optimal bias point, which suggests that even a simple single-photon detection is nearly optimal for this scheme. Details of the simulation and further discussions can be found in the Supplementary Material.

The exceptional performance of the example shown above is not accidental but rather a typical feature of our echoed random protocol. To quantitatively demonstrate this statistical robustness, we generated 1,000 distinct random probe states via numerical simulation and analyzed the resulting distribution of their metrological gains. As shown in Fig. \ref{fig2}(c), the metrological gain $G_\mathrm{c}(\theta_0)$ around the optimal bias point $\theta_\mathrm{b}$ exhibits a small standard deviation relative to the mean, indicating that most of the random probe states can achieve a similar enhancement. The inset of Fig. \ref{fig2}(c) shows that the distribution of $G_\mathrm{c,max}$ at the optimal bias point closely follows a Gaussian profile.

Beyond its statistical reliability, we now assess how the protocol's performance scales with the available resources. The scatter plot in Fig. \ref{fig2}(d) shows that as the maximum driving strength $\epsilon$ is increased, the average photon number $\ev{\hn}$ grows, which, in turn, is accompanied by an increase in the CFI. A power-law fit to the data yields $I_\mathrm{c,max}\propto  \ev{\hn}^{1.95}$, which indicates that despite the presence of measurement noise and the lack of probe state engineering, the metrological gain is still approaching the Heisenberg scaling $I_\mathrm{HL}\propto\ev{\hn}^2$. Note that for well-known states exhibiting Heisenberg scaling in single-mode phase sensing, such as squeezed vacuum states or maximum-variance states $\tfrac{1}{\sqrt{2}}(\ket{0}+\ket{N})$~\cite{Fadel2025}, current experimental techniques achieve high fidelity only for $\ev{\hn}<10$~\cite{Vahlbruch2016, Wang2019b, McCormick2019}, thereby limiting the attainable sensing precision. However, our echoed random metrology scheme is easily scalable to a high-dimensional Hilbert space by simply increasing the driving strength, as shown in Fig.~\ref{fig2}(d). For instance, with driving strength $\epsilon/\chi=400$, random probe states with $\ev{\hn}=100$ can be prepared.

We further explore the mean metrological gain over driving-pulse realizations as a function of the maximum driving strength $\epsilon$ and the nonlinear interaction resource $\chi T$, as shown in Fig.~\ref{fig2}(e). It can be seen that the metrological gain increases significantly with both the driving strength and nonlinear resources, indicating that $\chi T$ can serve as a metric for the quantum resources leveraged by the protocol. A metrological gain of nearly two orders of magnitude can be achieved around $\chi T\approx3$ and $\epsilon/\chi\approx 10^3$. These parameters are promisingly achievable in current superconducting systems~\cite{Iyama2024}. The analysis of the standard deviation in Fig.~\ref{fig2}(f) reveals that the system exhibits a transient regime at early times and enters a stabilization phase for $\chi T\gtrsim 1.5$ in the sense that the variance remains low. This implies that an increase in evolution time also enhances the system's statistical robustness. Specifically, in the upper-right region, the system's gain ($G_c \approx 90$) is much larger than its standard deviation ($\sigma_{G_c}\approx 10$).

Crucially, our protocol maintains high metrological gain over a broad range of pulse step sizes $\tau$. While the quantum Fisher information asymptotically vanishes in the extreme limit of $\tau \ll 1/(\epsilon^{2}T)$, as proven in the Supplementary Material, and diminishes as $\tau$ approaches the total evolution time $T$, it remains consistently large and statistically robust throughout the intermediate regime. This wide operational window ensures that the protocol does not require fine-tuning of the driving granularity to achieve significant gain. We emphasize that the numerical results presented here are not optimized over $\tau$, yet a clear metrological advantage is nonetheless observed.

In contrast to the operational flexibility of our protocol, traditional metrology schemes typically require the optimization of driving pulses to prepare sophisticated probe states, such as maximum-variance states $\frac{1}{\sqrt{2}}(\ket{0} + \ket{N})$. In large-scale quantum systems, achieving high-fidelity initial state preparation through quantum control methods like pulse optimization~\cite{Liu2017a} presents a significant challenge, which demands a large amount of classical computational resources~\cite{Lu2024a}. Random-pulse driving bypasses this complex optimization process and is therefore free from this limitation.

\begin{figure*}
    \centering
    \includegraphics[width=\linewidth]{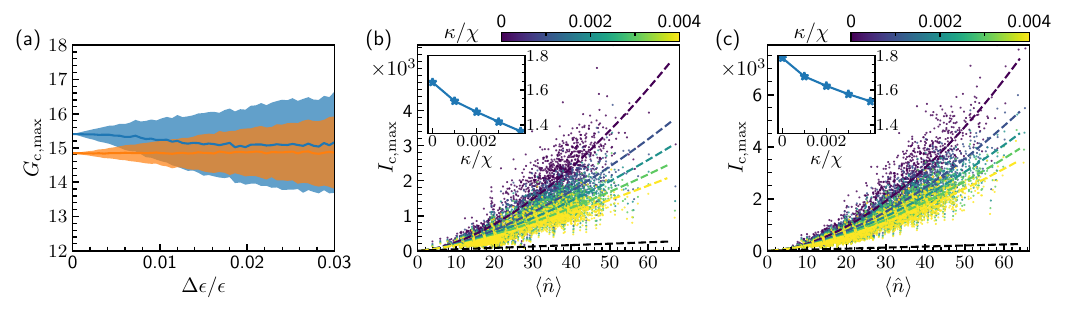}
    \caption{Robustness to control fluctuations and photon loss.
    (a) Metrological gain $G_{\mathrm{c,max}}$ versus relative fluctuation strength $\Delta\epsilon/\epsilon$ for random single-photon-driven (blue, $\ev{\hn}=34.8$) and two-photon-driven (orange, $\ev{\hn}=3.4$) probes. Parameters are $(\epsilon/\chi, \chi T) = (40, 1.5)$ and $(6, 0.8)$, respectively, with step size $\tau=0.02/\chi$ for both. Shaded regions indicate standard deviations over 1000 random realizations of fluctuations $\delta u_{1,2}(t)$.
    (b) Scatter plot of CFI $I_{\mathrm{c,max}}$ versus mean photon number $\ev{\hn}$ for single-photon drive under different photon-loss rates $\kappa$. The data are shown for $\chi T=1.5$ and various driving strengths $\epsilon/\chi \le 100$. Dashed curves are power-law fits to the data; the black dashed line marks the SQL $I_\mathrm{SQL}=4\ev{\hn}$. \textit{Inset:} power-law exponent $b$ from $I_\mathrm{c,max}=a\ev{\hn}^b$ as a function of $\kappa$.
    (c) Scatter plot of $I_\mathrm{c,max}$ versus $\ev{\hn}$ for two-photon drive under different photon-loss rates $\kappa$. The data are shown for $\chi T=0.8$ and various driving strengths $\epsilon/\chi \le 20$. Dashed curves are power-law fits to the data; the black dashed line marks the SQL $I_\mathrm{SQL}=4\ev{\hn}$. \textit{Inset:} power-law exponent $b$ from $I_\mathrm{c,max}=a\ev{\hn}^b$ as a function of $\kappa$.
    Other parameters: $\tau=0.1/\chi$, $\epsilon_\mathrm{dp}=10^{-3}$.
    }
    \label{fig3}
\end{figure*}

\textit{\textbf{Resilience to control fluctuations and loss}.-} We next examine how small variations in the control parameters affect the metrological gain. Specifically, we consider pulse-amplitude fluctuations~\cite{Egger2014} modeled as $\tilde{u}_{1,2}(t) = u_{1,2}(t) + \delta u_{1,2}(t)$, where $u_{1,2}(t)$ are the intended driving pulses and $\delta u_{1,2}(t)$ are zero-mean Gaussian fluctuations with a standard deviation $\Delta \epsilon$ that are uncorrelated in time. Figure~\ref{fig3}(a) shows the metrological gain under control fluctuations. The results show that the scheme is insensitive to small variations in the driving pulses, as the random nature of the driving pulses makes them largely unaffected by Gaussian fluctuations in their amplitudes.



Even though photon loss is present in practical state preparation processes, the metrological gain in our echoed random metrology scheme is sufficiently robust against it. This is fully demonstrated by the following numerical calculations.
In the presence of photon loss, the channel $\mathcal{E}_{\hat{H}}$ is governed by the following master equation:
\begin{equation}
    \dot{\rho} = -i[\hat{H},\rho] + \kappa[\ha\rho\ha^\dagger-\frac{1}{2}(\ha^\dagger\ha\rho + \rho\ha^\dagger\ha)].
\end{equation}
As shown in Fig. \ref{fig3}(b), the sensitivity compared to SQL remains significantly enhanced despite photon loss.
Although the scaling factor of the metrological gain decreases as the error rate increases, the randomly driven nonlinear system nonetheless surpasses the classical limit over a considerable range. For instance, even when the decay rate is as large as $\kappa/\chi=0.004$, the average CFI of random probe states still scales as $I_\mathrm{c,max}=7\times\ev{\hn}^{1.36}$, surpassing the SQL. By increasing the driving strength, probe states with larger average numbers of photons can be prepared, which further improves the attainable sensing precision.

Moreover, we notice that the parity symmetry of the system can be exploited to mitigate errors by employing a two-photon drive~\cite{Leghtas2015} with the Hamiltonian:
\begin{equation}\label{eq:hamiltonian_twoPhoton}
    \hat{H}_\mathrm{tp}(t) = \chi(\hadag\ha)^2 + u_{\mathrm{tp}}(t)\hat{a}^{\dagger2}+u_{\mathrm{tp}}^*(t)\hat{a}^{2},
\end{equation}
where $u_{\mathrm{tp}}(t)$ is the complex amplitude of random drive pulse. Crucially, this Hamiltonian commutes with the parity operator $\hat{\varPi} = (-1)^{\hadag\ha}$, confining unitary dynamics to the even-parity symmetry sector when starting from the vacuum state $\ket{0}$. While single-photon loss acts as a parity-flip operator, shunting the state into the orthogonal odd manifold, this dissipation can be identified by employing a three-outcome POVM $\{\hat{M}_0=\dyad{0}, \hat{M}_1=\dyad{1}, \hat{M}_2=I-\hat{M}_0-\hat{M}_1\}$. Unlike a simple binary measurement, resolving the $\ket{1}$ state allows the protocol to herald parity-flip errors. Analogous to the single-photon drive, the two-photon drive also exhibits robustness against control fluctuations, as shown in Fig.~\ref{fig3}(a). More importantly, Fig.~\ref{fig3}(c) indicates that it maintains a higher scaling factor than the single-photon drive in the presence of photon loss, indicating superior robustness against this specific noise channel.

\textit{\textbf{Conclusions}.-} We have proposed an echoed random process for quantum metrology that eliminates the need for both probe-state and measurement optimization, exhibiting high scalability to high-dimensional quantum systems. As a demonstration, we considered a self-Kerr oscillator driven by random coherent pulses for phase estimation, achieving substantial metrological gains approaching Heisenberg scaling. While not designed to outperform the absolute best optimized schemes, this scheme still offers a practical, hardware-efficient alternative, showing that such unengineered states combined with echo-based measurements can still achieve strong metrological performance over broad parameter ranges with high probability.  Furthermore, this scheme is also insensitive to control fluctuations and resilient to photon loss. Beyond bosonic platforms, it extends naturally to qubit systems with many-body interactions that generate metrologically useful entangled states~\cite{Shi2024a}. Experimentally feasible and adaptable to tasks such as displacement estimation, this method offers a practical, optimization-free route to quantum-enhanced metrology in interacting systems.

The source code of this work is available in the GitHub repository~\cite{Liu2026}.

\begin{acknowledgments}
This work was funded by the National Key Research and Development Program of China (2022YFA1204704), Quantum Science and Technology-National Science and Technology Major Project (Grant Nos.~2021ZD0300200, 2021ZD0301500, and 2021ZD0303200), the National Natural Science Foundation of China (NSFC) (T2325022, U23A2074, 92265210, 92365301, 92565301, 12547179, 12574539, and 12550006), the CAS Project for Young Scientists in Basic Research (YSBR-049),  the Fundamental Research Funds for the Central Universities, and USTC Research Funds of the Double First-Class Initiative. C.-L.Z. was also supported by the Shenzhen International Quantum Academy (Grant No. SIQA2024KFKT02). The numerical calculations were performed at the Supercomputing Center of USTC. This work was partially carried out at the USTC Center for Micro and Nanoscale Research and Fabrication.
\end{acknowledgments}


\begin{thebibliography}{49}%
\makeatletter
\providecommand \@ifxundefined [1]{%
 \@ifx{#1\undefined}
}%
\providecommand \@ifnum [1]{%
 \ifnum #1\expandafter \@firstoftwo
 \else \expandafter \@secondoftwo
 \fi
}%
\providecommand \@ifx [1]{%
 \ifx #1\expandafter \@firstoftwo
 \else \expandafter \@secondoftwo
 \fi
}%
\providecommand \natexlab [1]{#1}%
\providecommand \enquote  [1]{``#1''}%
\providecommand \bibnamefont  [1]{#1}%
\providecommand \bibfnamefont [1]{#1}%
\providecommand \citenamefont [1]{#1}%
\providecommand \href@noop [0]{\@secondoftwo}%
\providecommand \href [0]{\begingroup \@sanitize@url \@href}%
\providecommand \@href[1]{\@@startlink{#1}\@@href}%
\providecommand \@@href[1]{\endgroup#1\@@endlink}%
\providecommand \@sanitize@url [0]{\catcode `\\12\catcode `\$12\catcode
  `\&12\catcode `\#12\catcode `\^12\catcode `\_12\catcode `\%12\relax}%
\providecommand \@@startlink[1]{}%
\providecommand \@@endlink[0]{}%
\providecommand \url  [0]{\begingroup\@sanitize@url \@url }%
\providecommand \@url [1]{\endgroup\@href {#1}{\urlprefix }}%
\providecommand \urlprefix  [0]{URL }%
\providecommand \Eprint [0]{\href }%
\providecommand \doibase [0]{http://dx.doi.org/}%
\providecommand \selectlanguage [0]{\@gobble}%
\providecommand \bibinfo  [0]{\@secondoftwo}%
\providecommand \bibfield  [0]{\@secondoftwo}%
\providecommand \translation [1]{[#1]}%
\providecommand \BibitemOpen [0]{}%
\providecommand \bibitemStop [0]{}%
\providecommand \bibitemNoStop [0]{.\EOS\space}%
\providecommand \EOS [0]{\spacefactor3000\relax}%
\providecommand \BibitemShut  [1]{\csname bibitem#1\endcsname}%
\let\auto@bib@innerbib\@empty
\bibitem [{\citenamefont {Giovannetti}\ \emph {et~al.}(2011)\citenamefont
  {Giovannetti}, \citenamefont {Lloyd},\ and\ \citenamefont
  {Maccone}}]{Giovannetti2011}%
  \BibitemOpen
  \bibfield  {author} {\bibinfo {author} {\bibfnamefont {V.}~\bibnamefont
  {Giovannetti}}, \bibinfo {author} {\bibfnamefont {S.}~\bibnamefont {Lloyd}},
  \ and\ \bibinfo {author} {\bibfnamefont {L.}~\bibnamefont {Maccone}},\
  }\bibfield  {title} {\enquote {\bibinfo {title} {Advances in quantum
  metrology},}\ }\href {\doibase 10.1038/nphoton.2011.35} {\bibfield  {journal}
  {\bibinfo  {journal} {Nature Photonics}\ }\textbf {\bibinfo {volume} {5}},\
  \bibinfo {pages} {222} (\bibinfo {year} {2011})}\BibitemShut {NoStop}%
\bibitem [{\citenamefont {Degen}\ \emph {et~al.}(2017)\citenamefont {Degen},
  \citenamefont {Reinhard},\ and\ \citenamefont {Cappellaro}}]{Degen2017}%
  \BibitemOpen
  \bibfield  {author} {\bibinfo {author} {\bibfnamefont {C.~L.}\ \bibnamefont
  {Degen}}, \bibinfo {author} {\bibfnamefont {F.}~\bibnamefont {Reinhard}}, \
  and\ \bibinfo {author} {\bibfnamefont {P.}~\bibnamefont {Cappellaro}},\
  }\bibfield  {title} {\enquote {\bibinfo {title} {Quantum sensing},}\ }\href
  {\doibase 10.1103/RevModPhys.89.035002} {\bibfield  {journal} {\bibinfo
  {journal} {Rev. Mod. Phys.}\ }\textbf {\bibinfo {volume} {89}},\ \bibinfo
  {pages} {035002} (\bibinfo {year} {2017})}\BibitemShut {NoStop}%
\bibitem [{\citenamefont {Bao}\ \emph {et~al.}(2024)\citenamefont {Bao},
  \citenamefont {Xu}, \citenamefont {Song}, \citenamefont {Wang}, \citenamefont
  {Xiang}, \citenamefont {Zhu}, \citenamefont {Chen}, \citenamefont {Jin},
  \citenamefont {Zhu}, \citenamefont {Gao}, \citenamefont {Wu}, \citenamefont
  {Zhang}, \citenamefont {Wang}, \citenamefont {Zou}, \citenamefont {Tan},
  \citenamefont {Zhang}, \citenamefont {Cui}, \citenamefont {Shen},
  \citenamefont {Zhong}, \citenamefont {Li}, \citenamefont {Deng},
  \citenamefont {Zhang}, \citenamefont {Dong}, \citenamefont {Zhang},
  \citenamefont {Liu}, \citenamefont {Zhao}, \citenamefont {Hao}, \citenamefont
  {Li}, \citenamefont {Wang}, \citenamefont {Song}, \citenamefont {Guo},
  \citenamefont {Huang},\ and\ \citenamefont {Wang}}]{Bao2024}%
  \BibitemOpen
  \bibfield  {author} {\bibinfo {author} {\bibfnamefont {Z.}~\bibnamefont
  {Bao}}, \bibinfo {author} {\bibfnamefont {S.}~\bibnamefont {Xu}}, \bibinfo
  {author} {\bibfnamefont {Z.}~\bibnamefont {Song}}, \bibinfo {author}
  {\bibfnamefont {K.}~\bibnamefont {Wang}}, \bibinfo {author} {\bibfnamefont
  {L.}~\bibnamefont {Xiang}}, \bibinfo {author} {\bibfnamefont
  {Z.}~\bibnamefont {Zhu}}, \bibinfo {author} {\bibfnamefont {J.}~\bibnamefont
  {Chen}}, \bibinfo {author} {\bibfnamefont {F.}~\bibnamefont {Jin}}, \bibinfo
  {author} {\bibfnamefont {X.}~\bibnamefont {Zhu}}, \bibinfo {author}
  {\bibfnamefont {Y.}~\bibnamefont {Gao}}, \bibinfo {author} {\bibfnamefont
  {Y.}~\bibnamefont {Wu}}, \bibinfo {author} {\bibfnamefont {C.}~\bibnamefont
  {Zhang}}, \bibinfo {author} {\bibfnamefont {N.}~\bibnamefont {Wang}},
  \bibinfo {author} {\bibfnamefont {Y.}~\bibnamefont {Zou}}, \bibinfo {author}
  {\bibfnamefont {Z.}~\bibnamefont {Tan}}, \bibinfo {author} {\bibfnamefont
  {A.}~\bibnamefont {Zhang}}, \bibinfo {author} {\bibfnamefont
  {Z.}~\bibnamefont {Cui}}, \bibinfo {author} {\bibfnamefont {F.}~\bibnamefont
  {Shen}}, \bibinfo {author} {\bibfnamefont {J.}~\bibnamefont {Zhong}},
  \bibinfo {author} {\bibfnamefont {T.}~\bibnamefont {Li}}, \bibinfo {author}
  {\bibfnamefont {J.}~\bibnamefont {Deng}}, \bibinfo {author} {\bibfnamefont
  {X.}~\bibnamefont {Zhang}}, \bibinfo {author} {\bibfnamefont
  {H.}~\bibnamefont {Dong}}, \bibinfo {author} {\bibfnamefont {P.}~\bibnamefont
  {Zhang}}, \bibinfo {author} {\bibfnamefont {Y.-R.}\ \bibnamefont {Liu}},
  \bibinfo {author} {\bibfnamefont {L.}~\bibnamefont {Zhao}}, \bibinfo {author}
  {\bibfnamefont {J.}~\bibnamefont {Hao}}, \bibinfo {author} {\bibfnamefont
  {H.}~\bibnamefont {Li}}, \bibinfo {author} {\bibfnamefont {Z.}~\bibnamefont
  {Wang}}, \bibinfo {author} {\bibfnamefont {C.}~\bibnamefont {Song}}, \bibinfo
  {author} {\bibfnamefont {Q.}~\bibnamefont {Guo}}, \bibinfo {author}
  {\bibfnamefont {B.}~\bibnamefont {Huang}}, \ and\ \bibinfo {author}
  {\bibfnamefont {H.}~\bibnamefont {Wang}},\ }\bibfield  {title} {\enquote
  {\bibinfo {title} {Creating and controlling global
  {{Greenberger-Horne-Zeilinger}} entanglement on quantum processors},}\ }\href
  {\doibase 10.1038/s41467-024-53140-5} {\bibfield  {journal} {\bibinfo
  {journal} {Nature Communications}\ }\textbf {\bibinfo {volume} {15}},\
  \bibinfo {pages} {8823} (\bibinfo {year} {2024})}\BibitemShut {NoStop}%
\bibitem [{\citenamefont {Deng}\ \emph {et~al.}(2024)\citenamefont {Deng},
  \citenamefont {Li}, \citenamefont {Chen}, \citenamefont {Ni}, \citenamefont
  {Cai}, \citenamefont {Mai}, \citenamefont {Zhang}, \citenamefont {Zheng},
  \citenamefont {Yu}, \citenamefont {Zou}, \citenamefont {Liu}, \citenamefont
  {Yan}, \citenamefont {Xu},\ and\ \citenamefont {Yu}}]{deng2024}%
  \BibitemOpen
  \bibfield  {author} {\bibinfo {author} {\bibfnamefont {X.}~\bibnamefont
  {Deng}}, \bibinfo {author} {\bibfnamefont {S.}~\bibnamefont {Li}}, \bibinfo
  {author} {\bibfnamefont {Z.-J.}\ \bibnamefont {Chen}}, \bibinfo {author}
  {\bibfnamefont {Z.}~\bibnamefont {Ni}}, \bibinfo {author} {\bibfnamefont
  {Y.}~\bibnamefont {Cai}}, \bibinfo {author} {\bibfnamefont {J.}~\bibnamefont
  {Mai}}, \bibinfo {author} {\bibfnamefont {L.}~\bibnamefont {Zhang}}, \bibinfo
  {author} {\bibfnamefont {P.}~\bibnamefont {Zheng}}, \bibinfo {author}
  {\bibfnamefont {H.}~\bibnamefont {Yu}}, \bibinfo {author} {\bibfnamefont
  {C.-L.}\ \bibnamefont {Zou}}, \bibinfo {author} {\bibfnamefont
  {S.}~\bibnamefont {Liu}}, \bibinfo {author} {\bibfnamefont {F.}~\bibnamefont
  {Yan}}, \bibinfo {author} {\bibfnamefont {Y.}~\bibnamefont {Xu}}, \ and\
  \bibinfo {author} {\bibfnamefont {D.}~\bibnamefont {Yu}},\ }\bibfield
  {title} {\enquote {\bibinfo {title} {Quantum-enhanced metrology with large
  {Fock} states},}\ }\href {\doibase 10.1038/s41567-024-02619-5} {\bibfield
  {journal} {\bibinfo  {journal} {Nature Physics}\ }\textbf {\bibinfo {volume}
  {20}},\ \bibinfo {pages} {1874} (\bibinfo {year} {2024})}\BibitemShut
  {NoStop}%
\bibitem [{\citenamefont {Wolf}\ \emph {et~al.}(2019)\citenamefont {Wolf},
  \citenamefont {Shi}, \citenamefont {Heip}, \citenamefont {Gessner},
  \citenamefont {Pezz{\`e}}, \citenamefont {Smerzi}, \citenamefont {Schulte},
  \citenamefont {Hammerer},\ and\ \citenamefont {Schmidt}}]{wolf2019}%
  \BibitemOpen
  \bibfield  {author} {\bibinfo {author} {\bibfnamefont {F.}~\bibnamefont
  {Wolf}}, \bibinfo {author} {\bibfnamefont {C.}~\bibnamefont {Shi}}, \bibinfo
  {author} {\bibfnamefont {J.~C.}\ \bibnamefont {Heip}}, \bibinfo {author}
  {\bibfnamefont {M.}~\bibnamefont {Gessner}}, \bibinfo {author} {\bibfnamefont
  {L.}~\bibnamefont {Pezz{\`e}}}, \bibinfo {author} {\bibfnamefont
  {A.}~\bibnamefont {Smerzi}}, \bibinfo {author} {\bibfnamefont
  {M.}~\bibnamefont {Schulte}}, \bibinfo {author} {\bibfnamefont
  {K.}~\bibnamefont {Hammerer}}, \ and\ \bibinfo {author} {\bibfnamefont
  {P.~O.}\ \bibnamefont {Schmidt}},\ }\bibfield  {title} {\enquote {\bibinfo
  {title} {Motional {Fock} states for quantum-enhanced amplitude and phase
  measurements with trapped ions},}\ }\href {\doibase
  10.1038/s41467-019-10576-4} {\bibfield  {journal} {\bibinfo  {journal}
  {Nature Communications}\ }\textbf {\bibinfo {volume} {10}},\ \bibinfo {pages}
  {2929} (\bibinfo {year} {2019})}\BibitemShut {NoStop}%
\bibitem [{\citenamefont {Eckner}\ \emph {et~al.}(2023)\citenamefont {Eckner},
  \citenamefont {Darkwah~Oppong}, \citenamefont {Cao}, \citenamefont {Young},
  \citenamefont {Milner}, \citenamefont {Robinson}, \citenamefont {Ye},\ and\
  \citenamefont {Kaufman}}]{Eckner2023}%
  \BibitemOpen
  \bibfield  {author} {\bibinfo {author} {\bibfnamefont {W.~J.}\ \bibnamefont
  {Eckner}}, \bibinfo {author} {\bibfnamefont {N.}~\bibnamefont
  {Darkwah~Oppong}}, \bibinfo {author} {\bibfnamefont {A.}~\bibnamefont {Cao}},
  \bibinfo {author} {\bibfnamefont {A.~W.}\ \bibnamefont {Young}}, \bibinfo
  {author} {\bibfnamefont {W.~R.}\ \bibnamefont {Milner}}, \bibinfo {author}
  {\bibfnamefont {J.~M.}\ \bibnamefont {Robinson}}, \bibinfo {author}
  {\bibfnamefont {J.}~\bibnamefont {Ye}}, \ and\ \bibinfo {author}
  {\bibfnamefont {A.~M.}\ \bibnamefont {Kaufman}},\ }\bibfield  {title}
  {\enquote {\bibinfo {title} {Realizing spin squeezing with {{Rydberg}}
  interactions in an optical clock},}\ }\href {\doibase
  10.1038/s41586-023-06360-6} {\bibfield  {journal} {\bibinfo  {journal}
  {Nature}\ }\textbf {\bibinfo {volume} {621}},\ \bibinfo {pages} {734}
  (\bibinfo {year} {2023})}\BibitemShut {NoStop}%
\bibitem [{\citenamefont {Wang}\ \emph {et~al.}(2019)\citenamefont {Wang},
  \citenamefont {Wu}, \citenamefont {Ma}, \citenamefont {Cai}, \citenamefont
  {Hu}, \citenamefont {Mu}, \citenamefont {Xu}, \citenamefont {Chen},
  \citenamefont {Wang}, \citenamefont {Song}, \citenamefont {Yuan},
  \citenamefont {Zou}, \citenamefont {Duan},\ and\ \citenamefont
  {Sun}}]{Wang2019b}%
  \BibitemOpen
  \bibfield  {author} {\bibinfo {author} {\bibfnamefont {W.}~\bibnamefont
  {Wang}}, \bibinfo {author} {\bibfnamefont {Y.}~\bibnamefont {Wu}}, \bibinfo
  {author} {\bibfnamefont {Y.}~\bibnamefont {Ma}}, \bibinfo {author}
  {\bibfnamefont {W.}~\bibnamefont {Cai}}, \bibinfo {author} {\bibfnamefont
  {L.}~\bibnamefont {Hu}}, \bibinfo {author} {\bibfnamefont {X.}~\bibnamefont
  {Mu}}, \bibinfo {author} {\bibfnamefont {Y.}~\bibnamefont {Xu}}, \bibinfo
  {author} {\bibfnamefont {Z.-J.}\ \bibnamefont {Chen}}, \bibinfo {author}
  {\bibfnamefont {H.}~\bibnamefont {Wang}}, \bibinfo {author} {\bibfnamefont
  {Y.~P.}\ \bibnamefont {Song}}, \bibinfo {author} {\bibfnamefont
  {H.}~\bibnamefont {Yuan}}, \bibinfo {author} {\bibfnamefont {C.-L.}\
  \bibnamefont {Zou}}, \bibinfo {author} {\bibfnamefont {L.-M.}\ \bibnamefont
  {Duan}}, \ and\ \bibinfo {author} {\bibfnamefont {L.}~\bibnamefont {Sun}},\
  }\bibfield  {title} {\enquote {\bibinfo {title} {Heisenberg-limited
  single-mode quantum metrology in a superconducting circuit},}\ }\href
  {\doibase 10.1038/s41467-019-12290-7} {\bibfield  {journal} {\bibinfo
  {journal} {Nature Communications}\ }\textbf {\bibinfo {volume} {10}},\
  \bibinfo {pages} {4382} (\bibinfo {year} {2019})}\BibitemShut {NoStop}%
\bibitem [{\citenamefont {Pan}\ \emph {et~al.}(2025)\citenamefont {Pan},
  \citenamefont {Krisnanda}, \citenamefont {Duina}, \citenamefont {Park},
  \citenamefont {Song}, \citenamefont {Fontaine}, \citenamefont {Copetudo},
  \citenamefont {Filip},\ and\ \citenamefont {Gao}}]{Pan2025}%
  \BibitemOpen
  \bibfield  {author} {\bibinfo {author} {\bibfnamefont {X.}~\bibnamefont
  {Pan}}, \bibinfo {author} {\bibfnamefont {T.}~\bibnamefont {Krisnanda}},
  \bibinfo {author} {\bibfnamefont {A.}~\bibnamefont {Duina}}, \bibinfo
  {author} {\bibfnamefont {K.}~\bibnamefont {Park}}, \bibinfo {author}
  {\bibfnamefont {P.}~\bibnamefont {Song}}, \bibinfo {author} {\bibfnamefont
  {C.~Y.}\ \bibnamefont {Fontaine}}, \bibinfo {author} {\bibfnamefont
  {A.}~\bibnamefont {Copetudo}}, \bibinfo {author} {\bibfnamefont
  {R.}~\bibnamefont {Filip}}, \ and\ \bibinfo {author} {\bibfnamefont {Y.~Y.}\
  \bibnamefont {Gao}},\ }\bibfield  {title} {\enquote {\bibinfo {title}
  {Realization of versatile and effective quantum metrology using a single
  bosonic mode},}\ }\href {\doibase 10.1103/PRXQuantum.6.010304} {\bibfield
  {journal} {\bibinfo  {journal} {PRX Quantum}\ }\textbf {\bibinfo {volume}
  {6}},\ \bibinfo {pages} {010304} (\bibinfo {year} {2025})}\BibitemShut
  {NoStop}%
\bibitem [{\citenamefont {Mujal}\ \emph {et~al.}(2021)\citenamefont {Mujal},
  \citenamefont {{Mart{\'i}nez-Pe{\~n}a}}, \citenamefont {Nokkala},
  \citenamefont {{Garc{\'i}a-Beni}}, \citenamefont {Giorgi}, \citenamefont
  {Soriano},\ and\ \citenamefont {Zambrini}}]{Mujal2021}%
  \BibitemOpen
  \bibfield  {author} {\bibinfo {author} {\bibfnamefont {P.}~\bibnamefont
  {Mujal}}, \bibinfo {author} {\bibfnamefont {R.}~\bibnamefont
  {{Mart{\'i}nez-Pe{\~n}a}}}, \bibinfo {author} {\bibfnamefont
  {J.}~\bibnamefont {Nokkala}}, \bibinfo {author} {\bibfnamefont
  {J.}~\bibnamefont {{Garc{\'i}a-Beni}}}, \bibinfo {author} {\bibfnamefont
  {G.~L.}\ \bibnamefont {Giorgi}}, \bibinfo {author} {\bibfnamefont {M.~C.}\
  \bibnamefont {Soriano}}, \ and\ \bibinfo {author} {\bibfnamefont
  {R.}~\bibnamefont {Zambrini}},\ }\bibfield  {title} {\enquote {\bibinfo
  {title} {Opportunities in quantum reservoir computing and extreme learning
  machines},}\ }\href {\doibase 10.1002/qute.202100027} {\bibfield  {journal}
  {\bibinfo  {journal} {Advanced Quantum Technologies}\ }\textbf {\bibinfo
  {volume} {4}},\ \bibinfo {pages} {2100027} (\bibinfo {year}
  {2021})}\BibitemShut {NoStop}%
\bibitem [{\citenamefont {Hu}\ \emph {et~al.}(2024)\citenamefont {Hu},
  \citenamefont {Khan}, \citenamefont {Bronn}, \citenamefont {Angelatos},
  \citenamefont {Rowlands}, \citenamefont {Ribeill},\ and\ \citenamefont
  {T{\"u}reci}}]{Hu2024}%
  \BibitemOpen
  \bibfield  {author} {\bibinfo {author} {\bibfnamefont {F.}~\bibnamefont
  {Hu}}, \bibinfo {author} {\bibfnamefont {S.~A.}\ \bibnamefont {Khan}},
  \bibinfo {author} {\bibfnamefont {N.~T.}\ \bibnamefont {Bronn}}, \bibinfo
  {author} {\bibfnamefont {G.}~\bibnamefont {Angelatos}}, \bibinfo {author}
  {\bibfnamefont {G.~E.}\ \bibnamefont {Rowlands}}, \bibinfo {author}
  {\bibfnamefont {G.~J.}\ \bibnamefont {Ribeill}}, \ and\ \bibinfo {author}
  {\bibfnamefont {H.~E.}\ \bibnamefont {T{\"u}reci}},\ }\bibfield  {title}
  {\enquote {\bibinfo {title} {Overcoming the coherence time barrier in quantum
  machine learning on temporal data},}\ }\href {\doibase
  10.1038/s41467-024-51162-7} {\bibfield  {journal} {\bibinfo  {journal}
  {Nature Communications}\ }\textbf {\bibinfo {volume} {15}},\ \bibinfo {pages}
  {7491} (\bibinfo {year} {2024})}\BibitemShut {NoStop}%
\bibitem [{\citenamefont {Wang}\ \emph {et~al.}(2025)\citenamefont {Wang},
  \citenamefont {Sun}, \citenamefont {Kong}, \citenamefont {Sun},\ and\
  \citenamefont {Zhang}}]{Wang2025a}%
  \BibitemOpen
  \bibfield  {author} {\bibinfo {author} {\bibfnamefont {L.}~\bibnamefont
  {Wang}}, \bibinfo {author} {\bibfnamefont {P.}~\bibnamefont {Sun}}, \bibinfo
  {author} {\bibfnamefont {L.-J.}\ \bibnamefont {Kong}}, \bibinfo {author}
  {\bibfnamefont {Y.}~\bibnamefont {Sun}}, \ and\ \bibinfo {author}
  {\bibfnamefont {X.}~\bibnamefont {Zhang}},\ }\bibfield  {title} {\enquote
  {\bibinfo {title} {Quantum next-generation reservoir computing and its
  quantum optical implementation},}\ }\href {\doibase
  10.1103/PhysRevA.111.022609} {\bibfield  {journal} {\bibinfo  {journal}
  {Phys. Rev. A}\ }\textbf {\bibinfo {volume} {111}},\ \bibinfo {pages}
  {022609} (\bibinfo {year} {2025})}\BibitemShut {NoStop}%
\bibitem [{\citenamefont {Oszmaniec}\ \emph {et~al.}(2016)\citenamefont
  {Oszmaniec}, \citenamefont {Augusiak}, \citenamefont {Gogolin}, \citenamefont
  {{Ko{\l}ody{\'n} {\'n}ski}}, \citenamefont {Ac{\'{\i}}n},\ and\ \citenamefont
  {Lewenstein}}]{Oszmaniec2016}%
  \BibitemOpen
  \bibfield  {author} {\bibinfo {author} {\bibfnamefont {M.}~\bibnamefont
  {Oszmaniec}}, \bibinfo {author} {\bibfnamefont {R.}~\bibnamefont {Augusiak}},
  \bibinfo {author} {\bibfnamefont {C.}~\bibnamefont {Gogolin}}, \bibinfo
  {author} {\bibfnamefont {J.}~\bibnamefont {{Ko{\l}ody{\'n} {\'n}ski}}},
  \bibinfo {author} {\bibfnamefont {A.}~\bibnamefont {Ac{\'{\i}}n}}, \ and\
  \bibinfo {author} {\bibfnamefont {M.}~\bibnamefont {Lewenstein}},\ }\bibfield
   {title} {\enquote {\bibinfo {title} {Random bosonic states for robust
  quantum metrology},}\ }\href {\doibase 10.1103/PhysRevX.6.041044} {\bibfield
  {journal} {\bibinfo  {journal} {Phys. Rev. X}\ }\textbf {\bibinfo {volume}
  {6}},\ \bibinfo {pages} {041044} (\bibinfo {year} {2016})}\BibitemShut
  {NoStop}%
\bibitem [{\citenamefont {Shi}\ \emph {et~al.}(2025)\citenamefont {Shi},
  \citenamefont {Smerzi},\ and\ \citenamefont {Pezz{\`e}}}]{Shi2025}%
  \BibitemOpen
  \bibfield  {author} {\bibinfo {author} {\bibfnamefont {H.-L.}\ \bibnamefont
  {Shi}}, \bibinfo {author} {\bibfnamefont {A.}~\bibnamefont {Smerzi}}, \ and\
  \bibinfo {author} {\bibfnamefont {L.}~\bibnamefont {Pezz{\`e}}},\ }\bibfield
  {title} {\enquote {\bibinfo {title} {Quantum chaos, randomness and universal
  scaling of entanglement in various {{Krylov}} spaces},}\ }\href {\doibase
  10.21468/SciPostPhys.19.4.102} {\bibfield  {journal} {\bibinfo  {journal}
  {SciPost Phys.}\ }\textbf {\bibinfo {volume} {19}},\ \bibinfo {pages} {102}
  (\bibinfo {year} {2025})}\BibitemShut {NoStop}%
\bibitem [{\citenamefont {Kobrin}\ \emph {et~al.}(2024)\citenamefont {Kobrin},
  \citenamefont {Schuster}, \citenamefont {Block}, \citenamefont {Wu},
  \citenamefont {Mitchell}, \citenamefont {Davis},\ and\ \citenamefont
  {Yao}}]{Kobrin2024}%
  \BibitemOpen
  \bibfield  {author} {\bibinfo {author} {\bibfnamefont {B.}~\bibnamefont
  {Kobrin}}, \bibinfo {author} {\bibfnamefont {T.}~\bibnamefont {Schuster}},
  \bibinfo {author} {\bibfnamefont {M.}~\bibnamefont {Block}}, \bibinfo
  {author} {\bibfnamefont {W.}~\bibnamefont {Wu}}, \bibinfo {author}
  {\bibfnamefont {B.}~\bibnamefont {Mitchell}}, \bibinfo {author}
  {\bibfnamefont {E.}~\bibnamefont {Davis}}, \ and\ \bibinfo {author}
  {\bibfnamefont {N.~Y.}\ \bibnamefont {Yao}},\ }\href@noop {} {\enquote
  {\bibinfo {title} {A universal protocol for quantum-enhanced sensing via
  information scrambling},}\ } (\bibinfo {year} {2024})\BibitemShut {NoStop}%
\bibitem [{\citenamefont {Zhu}\ and\ \citenamefont {Hayashi}(2018)}]{Zhu2018}%
  \BibitemOpen
  \bibfield  {author} {\bibinfo {author} {\bibfnamefont {H.}~\bibnamefont
  {Zhu}}\ and\ \bibinfo {author} {\bibfnamefont {M.}~\bibnamefont {Hayashi}},\
  }\bibfield  {title} {\enquote {\bibinfo {title} {Universally fisher-symmetric
  informationally complete measurements},}\ }\href {\doibase
  10.1103/PhysRevLett.120.030404} {\bibfield  {journal} {\bibinfo  {journal}
  {Phys. Rev. Lett.}\ }\textbf {\bibinfo {volume} {120}},\ \bibinfo {pages}
  {030404} (\bibinfo {year} {2018})}\BibitemShut {NoStop}%
\bibitem [{\citenamefont {Hou}\ \emph {et~al.}(2018)\citenamefont {Hou},
  \citenamefont {Tang}, \citenamefont {Shang}, \citenamefont {Zhu},
  \citenamefont {Li}, \citenamefont {Yuan}, \citenamefont {Wu}, \citenamefont
  {Xiang}, \citenamefont {Li},\ and\ \citenamefont {Guo}}]{Hou2018}%
  \BibitemOpen
  \bibfield  {author} {\bibinfo {author} {\bibfnamefont {Z.}~\bibnamefont
  {Hou}}, \bibinfo {author} {\bibfnamefont {J.-F.}\ \bibnamefont {Tang}},
  \bibinfo {author} {\bibfnamefont {J.}~\bibnamefont {Shang}}, \bibinfo
  {author} {\bibfnamefont {H.}~\bibnamefont {Zhu}}, \bibinfo {author}
  {\bibfnamefont {J.}~\bibnamefont {Li}}, \bibinfo {author} {\bibfnamefont
  {Y.}~\bibnamefont {Yuan}}, \bibinfo {author} {\bibfnamefont {K.-D.}\
  \bibnamefont {Wu}}, \bibinfo {author} {\bibfnamefont {G.-Y.}\ \bibnamefont
  {Xiang}}, \bibinfo {author} {\bibfnamefont {C.-F.}\ \bibnamefont {Li}}, \
  and\ \bibinfo {author} {\bibfnamefont {G.-C.}\ \bibnamefont {Guo}},\
  }\bibfield  {title} {\enquote {\bibinfo {title} {Deterministic realization of
  collective measurements via photonic quantum walks},}\ }\href {\doibase
  10.1038/s41467-018-03849-x} {\bibfield  {journal} {\bibinfo  {journal}
  {Nature Communications}\ }\textbf {\bibinfo {volume} {9}},\ \bibinfo {pages}
  {1414} (\bibinfo {year} {2018})}\BibitemShut {NoStop}%
\bibitem [{\citenamefont {Zhou}\ and\ \citenamefont {Chen}(2025)}]{Zhou2025}%
  \BibitemOpen
  \bibfield  {author} {\bibinfo {author} {\bibfnamefont {S.}~\bibnamefont
  {Zhou}}\ and\ \bibinfo {author} {\bibfnamefont {S.}~\bibnamefont {Chen}},\
  }\href@noop {} {\enquote {\bibinfo {title} {Randomized measurements for
  multi-parameter quantum metrology},}\ } (\bibinfo {year} {2025})\BibitemShut
  {NoStop}%
\bibitem [{\citenamefont {Davis}\ \emph {et~al.}(2016)\citenamefont {Davis},
  \citenamefont {Bentsen},\ and\ \citenamefont {{Schleier-Smith}}}]{Davis2016}%
  \BibitemOpen
  \bibfield  {author} {\bibinfo {author} {\bibfnamefont {E.}~\bibnamefont
  {Davis}}, \bibinfo {author} {\bibfnamefont {G.}~\bibnamefont {Bentsen}}, \
  and\ \bibinfo {author} {\bibfnamefont {M.}~\bibnamefont {{Schleier-Smith}}},\
  }\bibfield  {title} {\enquote {\bibinfo {title} {Approaching the heisenberg
  limit without single-particle detection},}\ }\href {\doibase
  10.1103/PhysRevLett.116.053601} {\bibfield  {journal} {\bibinfo  {journal}
  {Phys. Rev. Lett.}\ }\textbf {\bibinfo {volume} {116}},\ \bibinfo {pages}
  {053601} (\bibinfo {year} {2016})}\BibitemShut {NoStop}%
\bibitem [{\citenamefont {Linnemann}\ \emph {et~al.}(2016)\citenamefont
  {Linnemann}, \citenamefont {Strobel}, \citenamefont {Muessel}, \citenamefont
  {Schulz}, \citenamefont {Lewis-Swan}, \citenamefont {Kheruntsyan},\ and\
  \citenamefont {Oberthaler}}]{Linnemann2016}%
  \BibitemOpen
  \bibfield  {author} {\bibinfo {author} {\bibfnamefont {D.}~\bibnamefont
  {Linnemann}}, \bibinfo {author} {\bibfnamefont {H.}~\bibnamefont {Strobel}},
  \bibinfo {author} {\bibfnamefont {W.}~\bibnamefont {Muessel}}, \bibinfo
  {author} {\bibfnamefont {J.}~\bibnamefont {Schulz}}, \bibinfo {author}
  {\bibfnamefont {R.~J.}\ \bibnamefont {Lewis-Swan}}, \bibinfo {author}
  {\bibfnamefont {K.~V.}\ \bibnamefont {Kheruntsyan}}, \ and\ \bibinfo {author}
  {\bibfnamefont {M.~K.}\ \bibnamefont {Oberthaler}},\ }\bibfield  {title}
  {\enquote {\bibinfo {title} {Quantum-enhanced sensing based on time reversal
  of nonlinear dynamics},}\ }\href {\doibase 10.1103/PhysRevLett.117.013001}
  {\bibfield  {journal} {\bibinfo  {journal} {Phys. Rev. Lett.}\ }\textbf
  {\bibinfo {volume} {117}},\ \bibinfo {pages} {013001} (\bibinfo {year}
  {2016})}\BibitemShut {NoStop}%
\bibitem [{\citenamefont {Nolan}\ \emph {et~al.}(2017)\citenamefont {Nolan},
  \citenamefont {Szigeti},\ and\ \citenamefont {Haine}}]{Nolan2017}%
  \BibitemOpen
  \bibfield  {author} {\bibinfo {author} {\bibfnamefont {S.~P.}\ \bibnamefont
  {Nolan}}, \bibinfo {author} {\bibfnamefont {S.~S.}\ \bibnamefont {Szigeti}},
  \ and\ \bibinfo {author} {\bibfnamefont {S.~A.}\ \bibnamefont {Haine}},\
  }\bibfield  {title} {\enquote {\bibinfo {title} {Optimal and robust quantum
  metrology using interaction-based readouts},}\ }\href {\doibase
  10.1103/PhysRevLett.119.193601} {\bibfield  {journal} {\bibinfo  {journal}
  {Phys. Rev. Lett.}\ }\textbf {\bibinfo {volume} {119}},\ \bibinfo {pages}
  {193601} (\bibinfo {year} {2017})}\BibitemShut {NoStop}%
\bibitem [{\citenamefont {Haine}(2018)}]{Haine2018}%
  \BibitemOpen
  \bibfield  {author} {\bibinfo {author} {\bibfnamefont {S.~A.}\ \bibnamefont
  {Haine}},\ }\bibfield  {title} {\enquote {\bibinfo {title} {Using
  interaction-based readouts to approach the ultimate limit of detection-noise
  robustness for quantum-enhanced metrology in collective spin systems},}\
  }\href {\doibase 10.1103/PhysRevA.98.030303} {\bibfield  {journal} {\bibinfo
  {journal} {Phys. Rev. A}\ }\textbf {\bibinfo {volume} {98}},\ \bibinfo
  {pages} {030303} (\bibinfo {year} {2018})}\BibitemShut {NoStop}%
\bibitem [{\citenamefont {Mirkhalaf}\ \emph {et~al.}(2018)\citenamefont
  {Mirkhalaf}, \citenamefont {Nolan},\ and\ \citenamefont
  {Haine}}]{Mirkhalaf2018}%
  \BibitemOpen
  \bibfield  {author} {\bibinfo {author} {\bibfnamefont {S.~S.}\ \bibnamefont
  {Mirkhalaf}}, \bibinfo {author} {\bibfnamefont {S.~P.}\ \bibnamefont
  {Nolan}}, \ and\ \bibinfo {author} {\bibfnamefont {S.~A.}\ \bibnamefont
  {Haine}},\ }\bibfield  {title} {\enquote {\bibinfo {title} {Robustifying
  twist-and-turn entanglement with interaction-based readout},}\ }\href
  {\doibase 10.1103/PhysRevA.97.053618} {\bibfield  {journal} {\bibinfo
  {journal} {Phys. Rev. A}\ }\textbf {\bibinfo {volume} {97}},\ \bibinfo
  {pages} {053618} (\bibinfo {year} {2018})}\BibitemShut {NoStop}%
\bibitem [{\citenamefont {Lewis-Swan}\ \emph {et~al.}(2020)\citenamefont
  {Lewis-Swan}, \citenamefont {Barberena}, \citenamefont {Muniz}, \citenamefont
  {Cline}, \citenamefont {Young}, \citenamefont {Thompson},\ and\ \citenamefont
  {Rey}}]{Lewis-Swan2020}%
  \BibitemOpen
  \bibfield  {author} {\bibinfo {author} {\bibfnamefont {R.~J.}\ \bibnamefont
  {Lewis-Swan}}, \bibinfo {author} {\bibfnamefont {D.}~\bibnamefont
  {Barberena}}, \bibinfo {author} {\bibfnamefont {J.~A.}\ \bibnamefont
  {Muniz}}, \bibinfo {author} {\bibfnamefont {J.~R.~K.}\ \bibnamefont {Cline}},
  \bibinfo {author} {\bibfnamefont {D.}~\bibnamefont {Young}}, \bibinfo
  {author} {\bibfnamefont {J.~K.}\ \bibnamefont {Thompson}}, \ and\ \bibinfo
  {author} {\bibfnamefont {A.~M.}\ \bibnamefont {Rey}},\ }\bibfield  {title}
  {\enquote {\bibinfo {title} {Protocol for precise field sensing in the
  optical domain with cold atoms in a cavity},}\ }\href {\doibase
  10.1103/PhysRevLett.124.193602} {\bibfield  {journal} {\bibinfo  {journal}
  {Phys. Rev. Lett.}\ }\textbf {\bibinfo {volume} {124}},\ \bibinfo {pages}
  {193602} (\bibinfo {year} {2020})}\BibitemShut {NoStop}%
\bibitem [{\citenamefont {Gilmore}\ \emph {et~al.}(2021)\citenamefont
  {Gilmore}, \citenamefont {Affolter}, \citenamefont {Lewis-Swan},
  \citenamefont {Barberena}, \citenamefont {Jordan}, \citenamefont {Rey},\ and\
  \citenamefont {Bollinger}}]{Gilmore2021}%
  \BibitemOpen
  \bibfield  {author} {\bibinfo {author} {\bibfnamefont {K.~A.}\ \bibnamefont
  {Gilmore}}, \bibinfo {author} {\bibfnamefont {M.}~\bibnamefont {Affolter}},
  \bibinfo {author} {\bibfnamefont {R.~J.}\ \bibnamefont {Lewis-Swan}},
  \bibinfo {author} {\bibfnamefont {D.}~\bibnamefont {Barberena}}, \bibinfo
  {author} {\bibfnamefont {E.}~\bibnamefont {Jordan}}, \bibinfo {author}
  {\bibfnamefont {A.~M.}\ \bibnamefont {Rey}}, \ and\ \bibinfo {author}
  {\bibfnamefont {J.~J.}\ \bibnamefont {Bollinger}},\ }\bibfield  {title}
  {\enquote {\bibinfo {title} {Quantum-enhanced sensing of displacements and
  electric fields with two-dimensional trapped-ion crystals},}\ }\href
  {\doibase 10.1126/science.abi5226} {\bibfield  {journal} {\bibinfo  {journal}
  {Science}\ }\textbf {\bibinfo {volume} {373}},\ \bibinfo {pages} {673}
  (\bibinfo {year} {2021})}\BibitemShut {NoStop}%
\bibitem [{\citenamefont {Colombo}\ \emph {et~al.}(2022)\citenamefont
  {Colombo}, \citenamefont {{Pedrozo-Pe{\~n}afiel}}, \citenamefont
  {Adiyatullin}, \citenamefont {Li}, \citenamefont {Mendez}, \citenamefont
  {Shu},\ and\ \citenamefont {Vuleti{\'c}}}]{Colombo2022}%
  \BibitemOpen
  \bibfield  {author} {\bibinfo {author} {\bibfnamefont {S.}~\bibnamefont
  {Colombo}}, \bibinfo {author} {\bibfnamefont {E.}~\bibnamefont
  {{Pedrozo-Pe{\~n}afiel}}}, \bibinfo {author} {\bibfnamefont {A.~F.}\
  \bibnamefont {Adiyatullin}}, \bibinfo {author} {\bibfnamefont
  {Z.}~\bibnamefont {Li}}, \bibinfo {author} {\bibfnamefont {E.}~\bibnamefont
  {Mendez}}, \bibinfo {author} {\bibfnamefont {C.}~\bibnamefont {Shu}}, \ and\
  \bibinfo {author} {\bibfnamefont {V.}~\bibnamefont {Vuleti{\'c}}},\
  }\bibfield  {title} {\enquote {\bibinfo {title} {Time-reversal-based quantum
  metrology with many-body entangled states},}\ }\href {\doibase
  10.1038/s41567-022-01653-5} {\bibfield  {journal} {\bibinfo  {journal}
  {Nature Physics}\ }\textbf {\bibinfo {volume} {18}},\ \bibinfo {pages} {925}
  (\bibinfo {year} {2022})}\BibitemShut {NoStop}%
\bibitem [{\citenamefont {Li}\ \emph {et~al.}(2023)\citenamefont {Li},
  \citenamefont {Colombo}, \citenamefont {Shu}, \citenamefont {Velez},
  \citenamefont {{Pilatowsky-Cameo}}, \citenamefont {Schmied}, \citenamefont
  {Choi}, \citenamefont {Lukin}, \citenamefont {{Pedrozo-Pe{\~n}afiel}},\ and\
  \citenamefont {Vuleti{\'c}}}]{Li2023a}%
  \BibitemOpen
  \bibfield  {author} {\bibinfo {author} {\bibfnamefont {Z.}~\bibnamefont
  {Li}}, \bibinfo {author} {\bibfnamefont {S.}~\bibnamefont {Colombo}},
  \bibinfo {author} {\bibfnamefont {C.}~\bibnamefont {Shu}}, \bibinfo {author}
  {\bibfnamefont {G.}~\bibnamefont {Velez}}, \bibinfo {author} {\bibfnamefont
  {S.}~\bibnamefont {{Pilatowsky-Cameo}}}, \bibinfo {author} {\bibfnamefont
  {R.}~\bibnamefont {Schmied}}, \bibinfo {author} {\bibfnamefont
  {S.}~\bibnamefont {Choi}}, \bibinfo {author} {\bibfnamefont {M.}~\bibnamefont
  {Lukin}}, \bibinfo {author} {\bibfnamefont {E.}~\bibnamefont
  {{Pedrozo-Pe{\~n}afiel}}}, \ and\ \bibinfo {author} {\bibfnamefont
  {V.}~\bibnamefont {Vuleti{\'c}}},\ }\bibfield  {title} {\enquote {\bibinfo
  {title} {Improving metrology with quantum scrambling},}\ }\href {\doibase
  10.1126/science.adg9500} {\bibfield  {journal} {\bibinfo  {journal}
  {Science}\ }\textbf {\bibinfo {volume} {380}},\ \bibinfo {pages} {1381}
  (\bibinfo {year} {2023})}\BibitemShut {NoStop}%
\bibitem [{\citenamefont {Guo}\ \emph {et~al.}(2024)\citenamefont {Guo},
  \citenamefont {He},\ and\ \citenamefont {Fadel}}]{Guo2024}%
  \BibitemOpen
  \bibfield  {author} {\bibinfo {author} {\bibfnamefont {J.}~\bibnamefont
  {Guo}}, \bibinfo {author} {\bibfnamefont {Q.}~\bibnamefont {He}}, \ and\
  \bibinfo {author} {\bibfnamefont {M.}~\bibnamefont {Fadel}},\ }\bibfield
  {title} {\enquote {\bibinfo {title} {Quantum metrology with a squeezed {Kerr}
  oscillator},}\ }\href {\doibase 10.1103/PhysRevA.109.052604} {\bibfield
  {journal} {\bibinfo  {journal} {Phys. Rev. A}\ }\textbf {\bibinfo {volume}
  {109}},\ \bibinfo {pages} {052604} (\bibinfo {year} {2024})}\BibitemShut
  {NoStop}%
\bibitem [{\citenamefont {{Google Quantum AI and
  Collaborators}}(2025)}]{Abanin2025}%
  \BibitemOpen
  \bibfield  {author} {\bibinfo {author} {\bibnamefont {{Google Quantum AI and
  Collaborators}}},\ }\bibfield  {title} {\enquote {\bibinfo {title}
  {Observation of constructive interference at the edge of quantum
  ergodicity},}\ }\href {\doibase 10.1038/s41586-025-09526-6} {\bibfield
  {journal} {\bibinfo  {journal} {Nature}\ }\textbf {\bibinfo {volume} {646}},\
  \bibinfo {pages} {825} (\bibinfo {year} {2025})}\BibitemShut {NoStop}%
\bibitem [{\citenamefont {Zurek}(2001)}]{Zurek2001}%
  \BibitemOpen
  \bibfield  {author} {\bibinfo {author} {\bibfnamefont {W.~H.}\ \bibnamefont
  {Zurek}},\ }\bibfield  {title} {\enquote {\bibinfo {title} {Sub-{Planck}
  structure in phase space and its relevance for quantum decoherence},}\ }\href
  {\doibase 10.1038/35089017} {\bibfield  {journal} {\bibinfo  {journal}
  {Nature}\ }\textbf {\bibinfo {volume} {412}},\ \bibinfo {pages} {712}
  (\bibinfo {year} {2001})}\BibitemShut {NoStop}%
\bibitem [{\citenamefont {Toscano}\ \emph {et~al.}(2006)\citenamefont
  {Toscano}, \citenamefont {Dalvit}, \citenamefont {Davidovich},\ and\
  \citenamefont {Zurek}}]{Toscano2006}%
  \BibitemOpen
  \bibfield  {author} {\bibinfo {author} {\bibfnamefont {F.}~\bibnamefont
  {Toscano}}, \bibinfo {author} {\bibfnamefont {D.~A.~R.}\ \bibnamefont
  {Dalvit}}, \bibinfo {author} {\bibfnamefont {L.}~\bibnamefont {Davidovich}},
  \ and\ \bibinfo {author} {\bibfnamefont {W.~H.}\ \bibnamefont {Zurek}},\
  }\bibfield  {title} {\enquote {\bibinfo {title} {Sub-{{Planck}} phase-space
  structures and {{Heisenberg-limited}} measurements},}\ }\href {\doibase
  10.1103/PhysRevA.73.023803} {\bibfield  {journal} {\bibinfo  {journal} {Phys.
  Rev. A}\ }\textbf {\bibinfo {volume} {73}},\ \bibinfo {pages} {023803}
  (\bibinfo {year} {2006})}\BibitemShut {NoStop}%
\bibitem [{\citenamefont {Wang}\ \emph {et~al.}(2022)\citenamefont {Wang},
  \citenamefont {Chen}, \citenamefont {Liu}, \citenamefont {Cai}, \citenamefont
  {Ma}, \citenamefont {Mu}, \citenamefont {Pan}, \citenamefont {Hua},
  \citenamefont {Hu}, \citenamefont {Xu}, \citenamefont {Wang}, \citenamefont
  {Song}, \citenamefont {Zou}, \citenamefont {Zou},\ and\ \citenamefont
  {Sun}}]{Wang2022a}%
  \BibitemOpen
  \bibfield  {author} {\bibinfo {author} {\bibfnamefont {W.}~\bibnamefont
  {Wang}}, \bibinfo {author} {\bibfnamefont {Z.-J.}\ \bibnamefont {Chen}},
  \bibinfo {author} {\bibfnamefont {X.}~\bibnamefont {Liu}}, \bibinfo {author}
  {\bibfnamefont {W.}~\bibnamefont {Cai}}, \bibinfo {author} {\bibfnamefont
  {Y.}~\bibnamefont {Ma}}, \bibinfo {author} {\bibfnamefont {X.}~\bibnamefont
  {Mu}}, \bibinfo {author} {\bibfnamefont {X.}~\bibnamefont {Pan}}, \bibinfo
  {author} {\bibfnamefont {Z.}~\bibnamefont {Hua}}, \bibinfo {author}
  {\bibfnamefont {L.}~\bibnamefont {Hu}}, \bibinfo {author} {\bibfnamefont
  {Y.}~\bibnamefont {Xu}}, \bibinfo {author} {\bibfnamefont {H.}~\bibnamefont
  {Wang}}, \bibinfo {author} {\bibfnamefont {Y.~P.}\ \bibnamefont {Song}},
  \bibinfo {author} {\bibfnamefont {X.-B.}\ \bibnamefont {Zou}}, \bibinfo
  {author} {\bibfnamefont {C.-L.}\ \bibnamefont {Zou}}, \ and\ \bibinfo
  {author} {\bibfnamefont {L.}~\bibnamefont {Sun}},\ }\bibfield  {title}
  {\enquote {\bibinfo {title} {Quantum-enhanced radiometry via approximate
  quantum error correction},}\ }\href {\doibase 10.1038/s41467-022-30410-8}
  {\bibfield  {journal} {\bibinfo  {journal} {Nature Communications}\ }\textbf
  {\bibinfo {volume} {13}},\ \bibinfo {pages} {3214} (\bibinfo {year}
  {2022})}\BibitemShut {NoStop}%
\bibitem [{\citenamefont {Nikolova}\ \emph {et~al.}(2025)\citenamefont
  {Nikolova}, \citenamefont {G\"ob}, \citenamefont {Singer},\ and\
  \citenamefont {Ivanov}}]{Nikolova2025}%
  \BibitemOpen
  \bibfield  {author} {\bibinfo {author} {\bibfnamefont {B.~S.}\ \bibnamefont
  {Nikolova}}, \bibinfo {author} {\bibfnamefont {M.}~\bibnamefont {G\"ob}},
  \bibinfo {author} {\bibfnamefont {K.}~\bibnamefont {Singer}}, \ and\ \bibinfo
  {author} {\bibfnamefont {P.~A.}\ \bibnamefont {Ivanov}},\ }\bibfield  {title}
  {\enquote {\bibinfo {title} {Mechanical squeezed kerr oscillator based on a
  tapered ion trap},}\ }\href {\doibase 10.1103/ybm3-72hs} {\bibfield
  {journal} {\bibinfo  {journal} {Phys. Rev. A}\ }\textbf {\bibinfo {volume}
  {112}},\ \bibinfo {pages} {013511} (\bibinfo {year} {2025})}\BibitemShut
  {NoStop}%
\bibitem [{\citenamefont {He}\ \emph {et~al.}(2023)\citenamefont {He},
  \citenamefont {Lu}, \citenamefont {Bao}, \citenamefont {Xue}, \citenamefont
  {Jiang}, \citenamefont {Wang}, \citenamefont {Roudsari}, \citenamefont
  {Delsing}, \citenamefont {Tsai},\ and\ \citenamefont {Lin}}]{He2023}%
  \BibitemOpen
  \bibfield  {author} {\bibinfo {author} {\bibfnamefont {X.~L.}\ \bibnamefont
  {He}}, \bibinfo {author} {\bibfnamefont {Y.}~\bibnamefont {Lu}}, \bibinfo
  {author} {\bibfnamefont {D.~Q.}\ \bibnamefont {Bao}}, \bibinfo {author}
  {\bibfnamefont {H.}~\bibnamefont {Xue}}, \bibinfo {author} {\bibfnamefont
  {W.~B.}\ \bibnamefont {Jiang}}, \bibinfo {author} {\bibfnamefont
  {Z.}~\bibnamefont {Wang}}, \bibinfo {author} {\bibfnamefont {A.~F.}\
  \bibnamefont {Roudsari}}, \bibinfo {author} {\bibfnamefont {P.}~\bibnamefont
  {Delsing}}, \bibinfo {author} {\bibfnamefont {J.~S.}\ \bibnamefont {Tsai}}, \
  and\ \bibinfo {author} {\bibfnamefont {Z.~R.}\ \bibnamefont {Lin}},\
  }\bibfield  {title} {\enquote {\bibinfo {title} {Fast generation of
  {{Schr{\"o}dinger}} cat states using a {{Kerr-tunable}} superconducting
  resonator},}\ }\href {\doibase 10.1038/s41467-023-42057-0} {\bibfield
  {journal} {\bibinfo  {journal} {Nature Communications}\ }\textbf {\bibinfo
  {volume} {14}},\ \bibinfo {pages} {6358} (\bibinfo {year}
  {2023})}\BibitemShut {NoStop}%
\bibitem [{\citenamefont {Xu}\ \emph {et~al.}(2025)\citenamefont {Xu},
  \citenamefont {Hua}, \citenamefont {Wang}, \citenamefont {Ma}, \citenamefont
  {Li}, \citenamefont {Chen}, \citenamefont {Zhou}, \citenamefont {Pan},
  \citenamefont {Xiao}, \citenamefont {Huang}, \citenamefont {Cai},
  \citenamefont {Ai}, \citenamefont {Liu}, \citenamefont {Zou},\ and\
  \citenamefont {Sun}}]{Xu2025}%
  \BibitemOpen
  \bibfield  {author} {\bibinfo {author} {\bibfnamefont {Y.}~\bibnamefont
  {Xu}}, \bibinfo {author} {\bibfnamefont {Z.}~\bibnamefont {Hua}}, \bibinfo
  {author} {\bibfnamefont {W.}~\bibnamefont {Wang}}, \bibinfo {author}
  {\bibfnamefont {Y.}~\bibnamefont {Ma}}, \bibinfo {author} {\bibfnamefont
  {M.}~\bibnamefont {Li}}, \bibinfo {author} {\bibfnamefont {J.}~\bibnamefont
  {Chen}}, \bibinfo {author} {\bibfnamefont {J.}~\bibnamefont {Zhou}}, \bibinfo
  {author} {\bibfnamefont {X.}~\bibnamefont {Pan}}, \bibinfo {author}
  {\bibfnamefont {L.}~\bibnamefont {Xiao}}, \bibinfo {author} {\bibfnamefont
  {H.}~\bibnamefont {Huang}}, \bibinfo {author} {\bibfnamefont
  {W.}~\bibnamefont {Cai}}, \bibinfo {author} {\bibfnamefont {H.}~\bibnamefont
  {Ai}}, \bibinfo {author} {\bibfnamefont {Y.-x.}\ \bibnamefont {Liu}},
  \bibinfo {author} {\bibfnamefont {C.-L.}\ \bibnamefont {Zou}}, \ and\
  \bibinfo {author} {\bibfnamefont {L.}~\bibnamefont {Sun}},\ }\bibfield
  {title} {\enquote {\bibinfo {title} {Dynamic compensation for pump-induced
  frequency shift in kerr-cat qubit initialization},}\ }\href {\doibase
  10.1103/PhysRevApplied.23.034060} {\bibfield  {journal} {\bibinfo  {journal}
  {Phys. Rev. Appl.}\ }\textbf {\bibinfo {volume} {23}},\ \bibinfo {pages}
  {034060} (\bibinfo {year} {2025})}\BibitemShut {NoStop}%
\bibitem [{\citenamefont {Yuan}\ and\ \citenamefont {Fung}(2017)}]{Yuan2017}%
  \BibitemOpen
  \bibfield  {author} {\bibinfo {author} {\bibfnamefont {H.}~\bibnamefont
  {Yuan}}\ and\ \bibinfo {author} {\bibfnamefont {C.-H.~F.}\ \bibnamefont
  {Fung}},\ }\bibfield  {title} {\enquote {\bibinfo {title} {Fidelity and
  fisher information on quantum channels},}\ }\href {\doibase
  10.1088/1367-2630/aa874c} {\bibfield  {journal} {\bibinfo  {journal} {New
  Journal of Physics}\ }\textbf {\bibinfo {volume} {19}},\ \bibinfo {pages}
  {113039} (\bibinfo {year} {2017})}\BibitemShut {NoStop}%
\bibitem [{\citenamefont {Heeres}\ \emph {et~al.}(2015)\citenamefont {Heeres},
  \citenamefont {Vlastakis}, \citenamefont {Holland}, \citenamefont
  {Krastanov}, \citenamefont {Albert}, \citenamefont {Frunzio}, \citenamefont
  {Jiang},\ and\ \citenamefont {Schoelkopf}}]{Heeres2015}%
  \BibitemOpen
  \bibfield  {author} {\bibinfo {author} {\bibfnamefont {R.~W.}\ \bibnamefont
  {Heeres}}, \bibinfo {author} {\bibfnamefont {B.}~\bibnamefont {Vlastakis}},
  \bibinfo {author} {\bibfnamefont {E.}~\bibnamefont {Holland}}, \bibinfo
  {author} {\bibfnamefont {S.}~\bibnamefont {Krastanov}}, \bibinfo {author}
  {\bibfnamefont {V.~V.}\ \bibnamefont {Albert}}, \bibinfo {author}
  {\bibfnamefont {L.}~\bibnamefont {Frunzio}}, \bibinfo {author} {\bibfnamefont
  {L.}~\bibnamefont {Jiang}}, \ and\ \bibinfo {author} {\bibfnamefont {R.~J.}\
  \bibnamefont {Schoelkopf}},\ }\bibfield  {title} {\enquote {\bibinfo {title}
  {Cavity state manipulation using photon-number selective phase gates},}\
  }\href {\doibase 10.1103/PhysRevLett.115.137002} {\bibfield  {journal}
  {\bibinfo  {journal} {Phys. Rev. Lett.}\ }\textbf {\bibinfo {volume} {115}},\
  \bibinfo {pages} {137002} (\bibinfo {year} {2015})}\BibitemShut {NoStop}%
\bibitem [{\citenamefont {Cai}\ \emph {et~al.}(2021)\citenamefont {Cai},
  \citenamefont {Ma}, \citenamefont {Wang}, \citenamefont {Zou},\ and\
  \citenamefont {Sun}}]{Cai2021}%
  \BibitemOpen
  \bibfield  {author} {\bibinfo {author} {\bibfnamefont {W.}~\bibnamefont
  {Cai}}, \bibinfo {author} {\bibfnamefont {Y.}~\bibnamefont {Ma}}, \bibinfo
  {author} {\bibfnamefont {W.}~\bibnamefont {Wang}}, \bibinfo {author}
  {\bibfnamefont {C.-L.}\ \bibnamefont {Zou}}, \ and\ \bibinfo {author}
  {\bibfnamefont {L.}~\bibnamefont {Sun}},\ }\bibfield  {title} {\enquote
  {\bibinfo {title} {Bosonic quantum error correction codes in superconducting
  quantum circuits},}\ }\href {\doibase
  https://doi.org/10.1016/j.fmre.2020.12.006} {\bibfield  {journal} {\bibinfo
  {journal} {Fundamental Research}\ }\textbf {\bibinfo {volume} {1}},\ \bibinfo
  {pages} {50} (\bibinfo {year} {2021})}\BibitemShut {NoStop}%
\bibitem [{\citenamefont {Meekhof}\ \emph {et~al.}(1996)\citenamefont
  {Meekhof}, \citenamefont {Monroe}, \citenamefont {King}, \citenamefont
  {Itano},\ and\ \citenamefont {Wineland}}]{Meekhof1996}%
  \BibitemOpen
  \bibfield  {author} {\bibinfo {author} {\bibfnamefont {D.~M.}\ \bibnamefont
  {Meekhof}}, \bibinfo {author} {\bibfnamefont {C.}~\bibnamefont {Monroe}},
  \bibinfo {author} {\bibfnamefont {B.~E.}\ \bibnamefont {King}}, \bibinfo
  {author} {\bibfnamefont {W.~M.}\ \bibnamefont {Itano}}, \ and\ \bibinfo
  {author} {\bibfnamefont {D.~J.}\ \bibnamefont {Wineland}},\ }\bibfield
  {title} {\enquote {\bibinfo {title} {Generation of nonclassical motional
  states of a trapped atom},}\ }\href {\doibase 10.1103/PhysRevLett.76.1796}
  {\bibfield  {journal} {\bibinfo  {journal} {Phys. Rev. Lett.}\ }\textbf
  {\bibinfo {volume} {76}},\ \bibinfo {pages} {1796} (\bibinfo {year}
  {1996})}\BibitemShut {NoStop}%
\bibitem [{\citenamefont {Li}\ \emph {et~al.}(2025)\citenamefont {Li},
  \citenamefont {Li}, \citenamefont {Cheng}, \citenamefont {Wang},
  \citenamefont {Zhao}, \citenamefont {Hou}, \citenamefont {Rehan},
  \citenamefont {Zhu}, \citenamefont {Yan}, \citenamefont {Qin}, \citenamefont
  {Peng}, \citenamefont {Yuan}, \citenamefont {Lin},\ and\ \citenamefont
  {Du}}]{Li2025}%
  \BibitemOpen
  \bibfield  {author} {\bibinfo {author} {\bibfnamefont {Y.}~\bibnamefont
  {Li}}, \bibinfo {author} {\bibfnamefont {Y.}~\bibnamefont {Li}}, \bibinfo
  {author} {\bibfnamefont {X.}~\bibnamefont {Cheng}}, \bibinfo {author}
  {\bibfnamefont {L.}~\bibnamefont {Wang}}, \bibinfo {author} {\bibfnamefont
  {X.}~\bibnamefont {Zhao}}, \bibinfo {author} {\bibfnamefont {W.}~\bibnamefont
  {Hou}}, \bibinfo {author} {\bibfnamefont {K.}~\bibnamefont {Rehan}}, \bibinfo
  {author} {\bibfnamefont {M.}~\bibnamefont {Zhu}}, \bibinfo {author}
  {\bibfnamefont {L.}~\bibnamefont {Yan}}, \bibinfo {author} {\bibfnamefont
  {X.}~\bibnamefont {Qin}}, \bibinfo {author} {\bibfnamefont {X.}~\bibnamefont
  {Peng}}, \bibinfo {author} {\bibfnamefont {H.}~\bibnamefont {Yuan}}, \bibinfo
  {author} {\bibfnamefont {Y.}~\bibnamefont {Lin}}, \ and\ \bibinfo {author}
  {\bibfnamefont {J.}~\bibnamefont {Du}},\ }\bibfield  {title} {\enquote
  {\bibinfo {title} {Programmable multi-mode entanglement via dissipative
  engineering in vibrating trapped ions},}\ }\href {\doibase
  10.1126/sciadv.adv7838} {\bibfield  {journal} {\bibinfo  {journal} {Science
  Advances}\ }\textbf {\bibinfo {volume} {11}},\ \bibinfo {pages} {eadv7838}
  (\bibinfo {year} {2025})}\BibitemShut {NoStop}%
\bibitem [{\citenamefont {Fadel}\ \emph {et~al.}(2025)\citenamefont {Fadel},
  \citenamefont {Roux},\ and\ \citenamefont {Gessner}}]{Fadel2025}%
  \BibitemOpen
  \bibfield  {author} {\bibinfo {author} {\bibfnamefont {M.}~\bibnamefont
  {Fadel}}, \bibinfo {author} {\bibfnamefont {N.}~\bibnamefont {Roux}}, \ and\
  \bibinfo {author} {\bibfnamefont {M.}~\bibnamefont {Gessner}},\ }\bibfield
  {title} {\enquote {\bibinfo {title} {Quantum metrology with a
  continuous-variable system},}\ }\href {\doibase 10.1088/1361-6633/ae00d8}
  {\bibfield  {journal} {\bibinfo  {journal} {Reports on Progress in Physics}\
  }\textbf {\bibinfo {volume} {88}},\ \bibinfo {pages} {106001} (\bibinfo
  {year} {2025})}\BibitemShut {NoStop}%
\bibitem [{\citenamefont {Vahlbruch}\ \emph {et~al.}(2016)\citenamefont
  {Vahlbruch}, \citenamefont {Mehmet}, \citenamefont {Danzmann},\ and\
  \citenamefont {Schnabel}}]{Vahlbruch2016}%
  \BibitemOpen
  \bibfield  {author} {\bibinfo {author} {\bibfnamefont {H.}~\bibnamefont
  {Vahlbruch}}, \bibinfo {author} {\bibfnamefont {M.}~\bibnamefont {Mehmet}},
  \bibinfo {author} {\bibfnamefont {K.}~\bibnamefont {Danzmann}}, \ and\
  \bibinfo {author} {\bibfnamefont {R.}~\bibnamefont {Schnabel}},\ }\bibfield
  {title} {\enquote {\bibinfo {title} {Detection of 15 {{dB}} squeezed states
  of light and their application for the absolute calibration of photoelectric
  quantum efficiency},}\ }\href {\doibase 10.1103/PhysRevLett.117.110801}
  {\bibfield  {journal} {\bibinfo  {journal} {Phys. Rev. Lett.}\ }\textbf
  {\bibinfo {volume} {117}},\ \bibinfo {pages} {110801} (\bibinfo {year}
  {2016})}\BibitemShut {NoStop}%
\bibitem [{\citenamefont {McCormick}\ \emph {et~al.}(2019)\citenamefont
  {McCormick}, \citenamefont {Keller}, \citenamefont {Burd}, \citenamefont
  {Wineland}, \citenamefont {Wilson},\ and\ \citenamefont
  {Leibfried}}]{McCormick2019}%
  \BibitemOpen
  \bibfield  {author} {\bibinfo {author} {\bibfnamefont {K.~C.}\ \bibnamefont
  {McCormick}}, \bibinfo {author} {\bibfnamefont {J.}~\bibnamefont {Keller}},
  \bibinfo {author} {\bibfnamefont {S.~C.}\ \bibnamefont {Burd}}, \bibinfo
  {author} {\bibfnamefont {D.~J.}\ \bibnamefont {Wineland}}, \bibinfo {author}
  {\bibfnamefont {A.~C.}\ \bibnamefont {Wilson}}, \ and\ \bibinfo {author}
  {\bibfnamefont {D.}~\bibnamefont {Leibfried}},\ }\bibfield  {title} {\enquote
  {\bibinfo {title} {Quantum-enhanced sensing of a single-ion mechanical
  oscillator},}\ }\href {\doibase 10.1038/s41586-019-1421-y} {\bibfield
  {journal} {\bibinfo  {journal} {Nature}\ }\textbf {\bibinfo {volume} {572}},\
  \bibinfo {pages} {86} (\bibinfo {year} {2019})}\BibitemShut {NoStop}%
\bibitem [{\citenamefont {Iyama}\ \emph {et~al.}(2024)\citenamefont {Iyama},
  \citenamefont {Kamiya}, \citenamefont {Fujii}, \citenamefont {Mukai},
  \citenamefont {Zhou}, \citenamefont {Nagase}, \citenamefont {Tomonaga},
  \citenamefont {Wang}, \citenamefont {Xue}, \citenamefont {Watabe},
  \citenamefont {Kwon},\ and\ \citenamefont {Tsai}}]{Iyama2024}%
  \BibitemOpen
  \bibfield  {author} {\bibinfo {author} {\bibfnamefont {D.}~\bibnamefont
  {Iyama}}, \bibinfo {author} {\bibfnamefont {T.}~\bibnamefont {Kamiya}},
  \bibinfo {author} {\bibfnamefont {S.}~\bibnamefont {Fujii}}, \bibinfo
  {author} {\bibfnamefont {H.}~\bibnamefont {Mukai}}, \bibinfo {author}
  {\bibfnamefont {Y.}~\bibnamefont {Zhou}}, \bibinfo {author} {\bibfnamefont
  {T.}~\bibnamefont {Nagase}}, \bibinfo {author} {\bibfnamefont
  {A.}~\bibnamefont {Tomonaga}}, \bibinfo {author} {\bibfnamefont
  {R.}~\bibnamefont {Wang}}, \bibinfo {author} {\bibfnamefont {J.-J.}\
  \bibnamefont {Xue}}, \bibinfo {author} {\bibfnamefont {S.}~\bibnamefont
  {Watabe}}, \bibinfo {author} {\bibfnamefont {S.}~\bibnamefont {Kwon}}, \ and\
  \bibinfo {author} {\bibfnamefont {J.-S.}\ \bibnamefont {Tsai}},\ }\bibfield
  {title} {\enquote {\bibinfo {title} {Observation and manipulation of quantum
  interference in a superconducting {{Kerr}} parametric oscillator},}\ }\href
  {\doibase 10.1038/s41467-023-44496-1} {\bibfield  {journal} {\bibinfo
  {journal} {Nature Communications}\ }\textbf {\bibinfo {volume} {15}},\
  \bibinfo {pages} {86} (\bibinfo {year} {2024})}\BibitemShut {NoStop}%
\bibitem [{\citenamefont {Liu}\ and\ \citenamefont {Yuan}(2017)}]{Liu2017a}%
  \BibitemOpen
  \bibfield  {author} {\bibinfo {author} {\bibfnamefont {J.}~\bibnamefont
  {Liu}}\ and\ \bibinfo {author} {\bibfnamefont {H.}~\bibnamefont {Yuan}},\
  }\bibfield  {title} {\enquote {\bibinfo {title} {Quantum parameter estimation
  with optimal control},}\ }\href {\doibase 10.1103/PhysRevA.96.012117}
  {\bibfield  {journal} {\bibinfo  {journal} {Phys. Rev. A}\ }\textbf {\bibinfo
  {volume} {96}},\ \bibinfo {pages} {012117} (\bibinfo {year}
  {2017})}\BibitemShut {NoStop}%
\bibitem [{\citenamefont {Lu}\ \emph {et~al.}(2024)\citenamefont {Lu},
  \citenamefont {Joshi}, \citenamefont {San~Dinh},\ and\ \citenamefont
  {Koch}}]{Lu2024a}%
  \BibitemOpen
  \bibfield  {author} {\bibinfo {author} {\bibfnamefont {Y.}~\bibnamefont
  {Lu}}, \bibinfo {author} {\bibfnamefont {S.}~\bibnamefont {Joshi}}, \bibinfo
  {author} {\bibfnamefont {V.}~\bibnamefont {San~Dinh}}, \ and\ \bibinfo
  {author} {\bibfnamefont {J.}~\bibnamefont {Koch}},\ }\bibfield  {title}
  {\enquote {\bibinfo {title} {Optimal control of large quantum systems:
  Assessing memory and runtime performance of {{GRAPE}}},}\ }\href {\doibase
  10.1088/2399-6528/ad22e5} {\bibfield  {journal} {\bibinfo  {journal} {Journal
  of Physics Communications}\ }\textbf {\bibinfo {volume} {8}},\ \bibinfo
  {pages} {025002} (\bibinfo {year} {2024})}\BibitemShut {NoStop}%
\bibitem [{\citenamefont {Egger}\ and\ \citenamefont
  {Wilhelm}(2014)}]{Egger2014}%
  \BibitemOpen
  \bibfield  {author} {\bibinfo {author} {\bibfnamefont {D.~J.}\ \bibnamefont
  {Egger}}\ and\ \bibinfo {author} {\bibfnamefont {F.~K.}\ \bibnamefont
  {Wilhelm}},\ }\bibfield  {title} {\enquote {\bibinfo {title} {Adaptive hybrid
  optimal quantum control for imprecisely characterized systems},}\ }\href
  {\doibase 10.1103/PhysRevLett.112.240503} {\bibfield  {journal} {\bibinfo
  {journal} {Phys. Rev. Lett.}\ }\textbf {\bibinfo {volume} {112}},\ \bibinfo
  {pages} {240503} (\bibinfo {year} {2014})}\BibitemShut {NoStop}%
\bibitem [{\citenamefont {Leghtas}\ \emph {et~al.}(2015)\citenamefont
  {Leghtas}, \citenamefont {Touzard}, \citenamefont {Pop}, \citenamefont {Kou},
  \citenamefont {Vlastakis}, \citenamefont {Petrenko}, \citenamefont {Sliwa},
  \citenamefont {Narla}, \citenamefont {Shankar}, \citenamefont {Hatridge},
  \citenamefont {Reagor}, \citenamefont {Frunzio}, \citenamefont {Schoelkopf},
  \citenamefont {Mirrahimi},\ and\ \citenamefont {Devoret}}]{Leghtas2015}%
  \BibitemOpen
  \bibfield  {author} {\bibinfo {author} {\bibfnamefont {Z.}~\bibnamefont
  {Leghtas}}, \bibinfo {author} {\bibfnamefont {S.}~\bibnamefont {Touzard}},
  \bibinfo {author} {\bibfnamefont {I.~M.}\ \bibnamefont {Pop}}, \bibinfo
  {author} {\bibfnamefont {A.}~\bibnamefont {Kou}}, \bibinfo {author}
  {\bibfnamefont {B.}~\bibnamefont {Vlastakis}}, \bibinfo {author}
  {\bibfnamefont {A.}~\bibnamefont {Petrenko}}, \bibinfo {author}
  {\bibfnamefont {K.~M.}\ \bibnamefont {Sliwa}}, \bibinfo {author}
  {\bibfnamefont {A.}~\bibnamefont {Narla}}, \bibinfo {author} {\bibfnamefont
  {S.}~\bibnamefont {Shankar}}, \bibinfo {author} {\bibfnamefont {M.~J.}\
  \bibnamefont {Hatridge}}, \bibinfo {author} {\bibfnamefont {M.}~\bibnamefont
  {Reagor}}, \bibinfo {author} {\bibfnamefont {L.}~\bibnamefont {Frunzio}},
  \bibinfo {author} {\bibfnamefont {R.~J.}\ \bibnamefont {Schoelkopf}},
  \bibinfo {author} {\bibfnamefont {M.}~\bibnamefont {Mirrahimi}}, \ and\
  \bibinfo {author} {\bibfnamefont {M.~H.}\ \bibnamefont {Devoret}},\
  }\bibfield  {title} {\enquote {\bibinfo {title} {Confining the state of light
  to a quantum manifold by engineered two-photon loss},}\ }\href {\doibase
  10.1126/science.aaa2085} {\bibfield  {journal} {\bibinfo  {journal}
  {Science}\ }\textbf {\bibinfo {volume} {347}},\ \bibinfo {pages} {853}
  (\bibinfo {year} {2015})}\BibitemShut {NoStop}%
\bibitem [{\citenamefont {Shi}\ \emph {et~al.}(2024)\citenamefont {Shi},
  \citenamefont {Guan},\ and\ \citenamefont {Yang}}]{Shi2024a}%
  \BibitemOpen
  \bibfield  {author} {\bibinfo {author} {\bibfnamefont {H.-L.}\ \bibnamefont
  {Shi}}, \bibinfo {author} {\bibfnamefont {X.-W.}\ \bibnamefont {Guan}}, \
  and\ \bibinfo {author} {\bibfnamefont {J.}~\bibnamefont {Yang}},\ }\bibfield
  {title} {\enquote {\bibinfo {title} {Universal shot-noise limit for quantum
  metrology with local hamiltonians},}\ }\href {\doibase
  10.1103/PhysRevLett.132.100803} {\bibfield  {journal} {\bibinfo  {journal}
  {Phys. Rev. Lett.}\ }\textbf {\bibinfo {volume} {132}},\ \bibinfo {pages}
  {100803} (\bibinfo {year} {2024})}\BibitemShut {NoStop}%
\bibitem [{\citenamefont {Liu}\ \emph {et~al.}(2026)\citenamefont {Liu},
  \citenamefont {Chen}, \citenamefont {Hua}, \citenamefont {Jie}, \citenamefont
  {Cai}, \citenamefont {Li}, \citenamefont {Sun}, \citenamefont {Zou},
  \citenamefont {Ren},\ and\ \citenamefont {Guo}}]{Liu2026}%
  \BibitemOpen
  \bibfield  {author} {\bibinfo {author} {\bibfnamefont {D.-S.}\ \bibnamefont
  {Liu}}, \bibinfo {author} {\bibfnamefont {Z.-J.}\ \bibnamefont {Chen}},
  \bibinfo {author} {\bibfnamefont {Z.}\ \bibnamefont {Hua}}, \bibinfo {author} {\bibfnamefont {Y.}\ \bibnamefont {Zhou}}, \bibinfo
  {author} {\bibfnamefont {Q.-X.}\ \bibnamefont {Jie}}, \bibinfo {author}
  {\bibfnamefont {W.}\ \bibnamefont {Cai}}, \bibinfo {author} {\bibfnamefont
  {M.}~\bibnamefont {Li}}, \bibinfo {author} {\bibfnamefont {L.}~\bibnamefont
  {Sun}}, \bibinfo {author} {\bibfnamefont {C.-L.}\ \bibnamefont {Zou}},
  \bibinfo {author} {\bibfnamefont {X.-F.}\ \bibnamefont {Ren}}, \ and\
  \bibinfo {author} {\bibfnamefont {G.-C.}\ \bibnamefont {Guo}},\ }\href@noop
  {} {\enquote {\bibinfo {title} {Echoed random quantum metrology},}\ }\bibinfo
  {howpublished}
  {\href{https://github.com/DS-Liu/Echoed-Random-Quantum-Metrology.git}{GitHub}}
  (\bibinfo {year} {2026})\BibitemShut {NoStop}%
\end{thebibliography}
\end{document}